\documentclass[aps,pra,floatfix,10pt,twocolumn,superscriptaddress,nofootinbib]{revtex4-2}

\usepackage[T1]{fontenc}
\usepackage[utf8]{inputenc}
\usepackage[english]{babel}
\usepackage{CJKutf8} 
\usepackage{microtype}
\usepackage{nicefrac}
\usepackage{soul}        
\setstcolor{red}
\setul{}{1.5pt}
\usepackage{amsmath}
\usepackage{amsfonts}
\usepackage{amssymb}
\usepackage{amstext}
\usepackage{amsthm}
\usepackage{amsbsy}
\usepackage{mathtools}
\usepackage{mathdots}
\usepackage{bm}
\usepackage{dsfont}
\usepackage{braket}
\usepackage{graphicx}
\usepackage{float}
\usepackage{subfloat}      
\usepackage[outdir=./]{epstopdf} 
\usepackage[notransparent]{svg}  
\usepackage{array}
\usepackage{multirow}
\usepackage{makecell}
\usepackage{booktabs}
\newcolumntype{C}{>{$}c<{$}}
\AtBeginDocument{
\heavyrulewidth=.08em
\lightrulewidth=.05em
\cmidrulewidth=.03em
\belowrulesep=.65ex
\belowbottomsep=0pt
\aboverulesep=.4ex
\abovetopsep=0pt
\cmidrulesep=\doublerulesep
\cmidrulekern=.5em
\defaultaddspace=.5em
\tabcolsep=7pt
}
\usepackage{xcolor}
\definecolor{emerald}{rgb}{0.07, 0.53, 0.03}

\usepackage[colorlinks=true,
            citecolor=blue,
            urlcolor=cyan]{hyperref}

\usepackage{nameref}
\usepackage{circuitikz}  
\usepackage{todonotes}  
\usepackage{verbatim}
\makeatletter
\let\latex@addcontentsline\addcontentsline
\renewcommand{\addcontentsline}[3]{}
\makeatother

\newenvironment{nalign}{
    \begin{equation}
    \begin{aligned}
}{
    \end{aligned}
    \end{equation}
    \ignorespacesafterend
}

\newcommand{\five}{Q2~}
\newcommand{\eight}{Q3~}
\newcommand{\sixteen}{Q1~}
\newcommand{\fivens}{Q2}
\newcommand{\eightns}{Q3}
\newcommand{\sixteenns}{Q1}

\newcommand{\wm}{\omega_\mathrm{m}}
\newcommand{\wps}{\omega_\mathrm{p}^{*}}
\newcommand{\wq}{\omega_\mathrm{q}}
\newcommand{\ww}{\omega_\mathrm{w}}

\newcommand{\dqb}{ {q} }
\newcommand{\fbop}{ {w} }
\newcommand{\fbfreq}{{\omega_\mathrm{w}}}
\newcommand{\fbdet}{{\Delta_\mathrm{w}}}
\newcommand{\qbfbdet}{{\Delta_\mathrm{qw}}}

\newcommand{\mop}{ {m} }
\newcommand{\mondet}{{\Delta_\mathrm{m}}}
\newcommand{\monfreq}{{\omega_\mathrm{m}}}
\newcommand{\qbmondet}{{\Delta_\mathrm{qm}}}

\newcommand{\gqb}{ g_\mathrm{qw}}
\newcommand{\gqa}{ g_\mathrm{qm}}

\begin{document}

\title{Multimode Strong-Coupling Processes in Circuit QED Lattices}
\author{Won Chan Lee}
\affiliation{Department of Physics, University of Maryland, College Park, MD 20742, USA}
\affiliation{Joint Quantum Institute, NIST/University of Maryland, College Park, Maryland 20742 USA}

\author{Ali Fahimniya}
\affiliation{Joint Quantum Institute, NIST/University of Maryland, College Park, Maryland 20742 USA}
\affiliation{Joint Center for Quantum Information and Computer Science, NIST/University of Maryland, College Park, Maryland 20742 USA}

\author{Kellen O'Brien}
\affiliation{Department of Physics, University of Maryland, College Park, MD 20742, USA}
\affiliation{Joint Quantum Institute, NIST/University of Maryland, College Park, Maryland 20742 USA}

\author{\begin{CJK}{UTF8}{gbsn}Yu-Xin Wang (王语馨)\end{CJK}}
\affiliation{Joint Center for Quantum Information and Computer Science, NIST/University of Maryland, College Park, Maryland 20742 USA}

\author{Alexandra~Behne}
\affiliation{Department of Physics, University of Maryland, College Park, MD 20742, USA}
\affiliation{Joint Quantum Institute, NIST/University of Maryland, College Park, Maryland 20742 USA}
\affiliation{Joint Center for Quantum Information and Computer Science, NIST/University of Maryland, College Park, Maryland 20742 USA}

\author{Maya Amouzegar}
\altaffiliation{Current address: Johns Hopkins Applied Physics Laboratory, Laurel, Maryland 20723, USA}
\affiliation{Department of Physics, University of Maryland, College Park, MD 20742, USA}
\affiliation{Joint Quantum Institute, NIST/University of Maryland, College Park, Maryland 20742 USA}

\author{Alexey V. Gorshkov}
\affiliation{Joint Quantum Institute, NIST/University of Maryland, College Park, Maryland 20742 USA}
\affiliation{Joint Center for Quantum Information and Computer Science, NIST/University of Maryland, College Park, Maryland 20742 USA}

\author{Alicia J.  Koll\'ar}
\affiliation{Department of Physics, University of Maryland, College Park, MD 20742, USA}
\affiliation{Joint Quantum Institute, NIST/University of Maryland, College Park, Maryland 20742 USA}
\affiliation{Maryland Quantum Materials Center, Department of Physics, University of Maryland, College Park, MD 20742, USA}

\preprint{APS/123-QED}

\date{\today}

\begin{abstract}
Circuit QED systems provide an ideal platform for exploring the strong-coupling regime of multimode cavity QED. Here we present two new phenomena from multimode strong coupling: a circuit Lagrangian analysis which captures beyond tight-binding effects of strong photon-photon coupling and experimental observation of strong wave-mixing resonances in the qubit response. Our circuit analysis reveals qualitatively new features such as emergent band gaps, lifted degeneracies, broadened flat bands, and frequency-dependent hopping. Within the multimode photon environment, strong qubit-photon coupling in turn gives rise to multiphoton processes involving multiple normal modes. We demonstrate a strong four-wave-mixing process involving excitation of a qubit and simultaneous frequency conversion between modes. Notably, this wave-mixing process is dominated by localized flat-band modes of the photonic lattice, which exhibit the strongest coupling to the transmon qubit.
\end{abstract}

\maketitle

Strongly coupled multimode cavity QED systems can realize effective spin–spin interactions \cite{Kollar2017supermode, Kollar:apparatus, Vaidya:2018fp, Guo:2019ej, Guo:2019fr, Gopalakrishnan09, Gopalakrishnan10, Gopalakrishnan:2011jx, Gopalakrishnan:2012cf} mediated by photonic bound states \cite{Douglas_2015,John_Wang,Calaj_2016}, enabling the study of quantum spin models \cite{SpinGlass, Sherrington:spinglass, Marsh:ising}. However, these systems also exhibit a breakdown of excitation-number-conserving interaction processes \cite{casanova:DSC, Braak:beyondJC, Bosman:ultrastrongcoupling, Niemczyk:ultrastrongcoupling, Forn:ultrastrongcoupling, Kuzmin:superstrong}, strong renormalization of modes \cite{meiser:superstrong, Yoshihara:ultrastrongcoupling, Bosman:ultrastrongcoupling, Niemczyk:ultrastrongcoupling, Forn:ultrastrongcoupling, Kuzmin:superstrong}, and non-Markovian decay \cite{Painter_Markov}. These phenomena require theoretical descriptions beyond simple tight-binding models of the photon modes~\cite{Kollar:2019linegraph, Koch:AnnPhysBerl, Koch:TRS_breaking, Houck_Nature_QS, Nunnenkamp:hopping, latticepaper}, Jaynes–Cummings–type light-matter interactions \cite{JCoriginalpaper, steckquantum}, and the Lindblad master equation formalism~\cite{Carmichael:LesHouches}. The Circuit QED platform is ideal for exploring such strong-coupling effects since it naturally exhibits both strong qubit–photon \cite{Blais:CavityQED, Koch:transmon, vrajitoarea2024} and photon–photon interactions \cite{Sundaresan, Liu, Koch:TRS_breaking, Koch:AnnPhysBerl}, as well as lithographic control over the mode structure and long coherence times.

Here we present two examples of qualitatively new phenomena due to multimode strong coupling. First, we show that strong photon-photon coupling in a resonator (cavity) array leads to beyond tight-binding effects which induce qualitative changes in the photonic modes. We explore these effects theoretically in the context of a superconducting resonator lattice and show that the terms neglected in a simple tight-binding description can open band gaps, lift degeneracies between band edges, broaden flat bands, and introduce frequency-dependent hopping. Second, we experimentally demonstrate strong resonances associated with the joint excitation of a transmon qubit~\cite{Koch:transmon} and an undriven lattice mode, which arise due to strong-coupling-mediated frequency conversion. In particular, we find that this wave-mixing process is dominated by frequency conversion into localized flat band modes that are not directly driven.

\vskip 0.1in

\textit{Multimode QED Device}.---We consider a capacitively-coupled superconducting coplanar-waveguide (CPW) lattice \cite{Houck_Nature_QS, Underwood:imaging, Houck:earlylattice, Koch:AnnPhysBerl}, with built-in transmon qubits \cite{latticepaper, mattias:lattice}. A CPW resonator is a long, flexible waveguide with multiple harmonic modes whose frequencies depend only on the length of the resonator. Therefore, these resonators can be bent into various shapes without affecting the modes, and these systems can realize exotic, even hyperbolic, geometries for microwave photons \cite{Kollar:2019hyperbolic, Kollar:2019linegraph}. Qubits incorporated in these systems experience a photon-mediated interaction that is determined by the effective geometry of the photon modes \cite{Sundaresan, latticepaper, Douglas_2015, bienias:Hyperbolic}.

\begin{figure*}[t]
    \centering
    \includegraphics[width=0.95\linewidth]{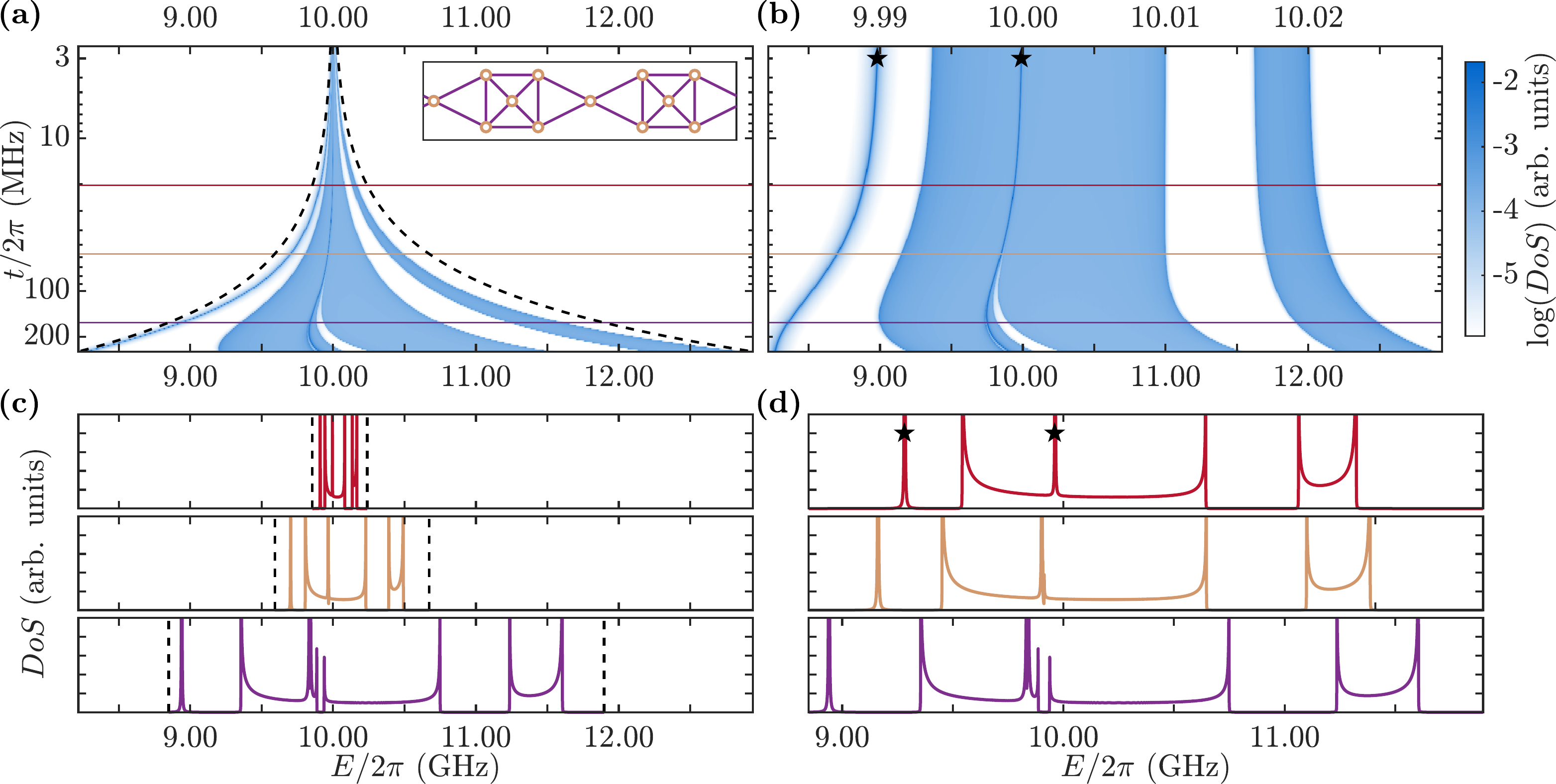}
    \caption{\label{fig:FW_tuning}
    \textit{Simulated Density of States of the Full-Wave Modes versus Nearest-Neighbor Hopping $t$.} 
    Density of states (DoS) as a function of energy $E$ and hopping $t$ (a) in a fixed energy window for all $t$ on a logarithmic scale and (b) in an energy window that scales linearly with $t$. The energy range used in (b) is indicated in (a) by dashed black lines. The inset in (a) shows the resonator connectivity of the device. (c)-(d) The line cuts of (a) and (b), respectively, are shown at three representative values of $t$ (colored horizontal lines in the corresponding 2D maps). Flat bands are marked by black stars in (b) and (d). In the simulations, we set $M = 100$.}
\end{figure*}

Our device, developed in Ref.~\cite{latticepaper}, is a quasi-1D lattice with nine unit cells, each composed of six resonators. Each resonator has the capacitor paddles of a transmon at one end, but only three of them include Josephson junctions, and therefore, form functional transmon qubits (see Appendix \ref{app:quasi-1Dlattice} for the device layout and the locations of the qubits). The inset of Fig.~\ref{fig:FW_tuning}(a) shows the resonator connectivity within two unit cells of the lattice. Resonators, represented by circles, are coupled to each other if a line connects them.

\vskip 0.1in

\textit{Beyond Tight-Binding Circuit Calculation}.---Previous theoretical treatments \cite{Kollar:2019linegraph, Koch:AnnPhysBerl, Koch:TRS_breaking, Houck_Nature_QS, Nunnenkamp:hopping, latticepaper} of the modes and band structures of CPW lattices considered a tight-binding approximation. However, these treatments neglect effects that are higher order in the bandwidth, such as the frequency-dependent impedance of coupling capacitors, intra-band scattering between harmonics of the CPW resonator, and counter-rotating coupling terms, which are present in the full circuit Lagrangian. These effects, while higher order, can give rise to qualitative changes in the photonic band structure at hopping strengths which are readily accessible in experiments. The perturbative approximation of Refs.~\cite{Kollar:2019linegraph, Koch:AnnPhysBerl, Koch:TRS_breaking, Houck_Nature_QS, Nunnenkamp:hopping, latticepaper} partially captures the effect of frequency-dependent coupling within a \emph{single} band. In this work, we present a systematic framework for calculating the modes of a CPW lattice and their couplings to transmon qubits in full, and show that the strong-coupling effects neglected by previous tight-binding treatments can fundamentally change the photonic density of states by opening band gaps, lifting degeneracies between band edges, broadening flat bands, and introducing frequency-dependent hopping.

We describe the CPW lattice using an effective lumped-element Lagrangian \cite{Painter_Markov, Painter_SSH} written in terms of the node fluxes and their time derivatives (the voltages) \cite{Blais:CircuitQED}, describing both the connections between CPW resonators and the on-site mode structure~\cite{Kollar:2019linegraph, Koch:AnnPhysBerl, Koch:TRS_breaking, Houck_Nature_QS, Nunnenkamp:hopping, latticepaper}. Absent couplings to other elements, each resonator exhibits intrinsic normal modes at frequencies $\omega_j=j\omega_0$, where $\omega_0$ is the fundamental frequency determined by the geometry of the waveguide, and $j$ is a positive integer. The flux profile of these normal modes is sinusoidal with $j$ half-wavelengths throughout the resonator and vanishing current at the ends. In the lumped-element circuit model~\cite{Goppl_2008}, the resonator is represented as a series of LC oscillators that reproduce these intrinsic normal modes. We introduce a UV cutoff by including only intrinsic normal modes with $j=1$ to $M$, where $M$ is finite, for each resonator. The total flux $\phi_{a}^{\rm tot}$ at the end of the resonator $a$ can be coupled to other elements (resonators or transmon qubits) with capacitors. This total flux is the sum of the individual oscillator fluxes $\phi_{a, j}$:
\begin{equation}
    \phi_a^{\rm tot} = \sum_{j=1}^M \phi_{a, j}.
\end{equation}
For simplicity, we consider a single coupling capacitor $C_c$ between resonators $a$ and $b$. 
The Lagrangian contribution for this coupler takes the form $\frac{1}{2} C_c (\dot\phi_a^{\rm tot} - \dot\phi_b^{\rm tot})^2$. Expanding this contribution leads to three kinds of terms: (i), $\frac{1}{2} C_c \dot \phi_{n,j}^2$ (where $n = a,b$) which renormalize the capacitors describing each individual mode; (ii), $-C_c \dot \phi_{a, j} \dot \phi_{b, j'}$ which provide either an intra-mode tight-binding coupling between the resonators when $j=j'$, or couple \textit{different} intrinsic modes of \textit{neighboring} resonators to each other; (iii), $C_c \dot \phi_{n, j} \dot \phi_{n, j'}$ (where $n=a,b$ and $j\neq j'$) which couple \textit{different} modes \textit{within} each resonator. The hopping parameter of the tight-binding coupling for the $j^{\text {th}}$ band is $t_j = jt$ where $t$ is proportional to $C_c$ and the renormalized fundamental resonator frequency~\cite{Koch:AnnPhysBerl, Koch:TRS_breaking, latticepaper}.
The analysis for the coupling of a resonator and a transmon qubit can be carried out similarly, leading to coupling between the qubit mode and the resonator’s intrinsic modes, as well as renormalization and multimode coupling among the resonator modes.

When the transmon modes are far detuned from the resonator modes, the quadratic Lagrangian with only resonator mode fluxes and their derivatives can be diagonalized to obtain the lattice normal modes. A solution for the CPW lattice [Eq.~\eqref{eq:res_lattice}] and a solution with the full transmon coupling description [Eq.~\eqref{eq:lattice_Ham}] can be found in Appendix \ref{app:bandcalculation}. Figure~\ref{fig:FW_tuning} shows the obtained density of states (DoS) as a result of this diagonalization for the full-wave (FW, $j=2$) band. The values of the circuit elements are chosen such that the on-site frequency of the FW modes, after renormalization due to the coupling capacitors, equals $2 \pi \times 10$ GHz. To illustrate the beyond tight-binding effects present in the full circuit calculation, the effective tight-binding hopping parameter $t$ is varied from $2\pi\times2.5$~MHz to $2\pi\times250$~MHz (we set $\hbar=1$ throughout this manuscript). The coupling capacitance between each resonator and its transmon qubit is chosen to be half the inter-resonator capacitance.

We analyze how the band structure evolves as a function of the hopping $t$. As $t$ increases, the bands not only broaden, but second-order effects absent in a tight-binding description also become apparent. Most notably, the bands initially expand linearly with $t$, as expected from a tight-binding model, as shown in the $t$-independent region in Fig.~\ref{fig:FW_tuning}(b). However, for larger $t$, the bands at higher energies become disproportionately wider, previously observed as \textit{frequency-dependent hopping} for this device~\cite{latticepaper}, and second-order band repulsion appears. The two flat bands in the tight-binding limit, denoted by black stars in Fig.~\ref{fig:FW_tuning}(b) and (d), remain flat as $t$ increases, except for a slight mixing of the flat band around $2\pi\times10$ GHz with dispersive bands. This flat band originally lies at a linear crossing of dispersive bands, but the capacitance of the transmons and their couplings to the resonators introduce additional mixing between the dispersive bands and open gaps at the nearby crossing (see Appendix Fig.~\ref{fig:band_structure_convergence} for details.).

\vskip 0.1in

\begin{figure}[t]
\centering
    \includegraphics[width=0.48\textwidth]{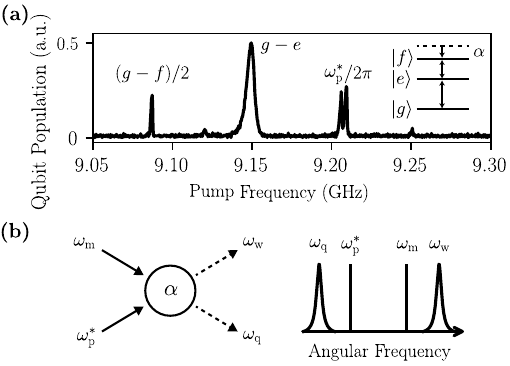}
    \vspace{-0.2cm}
    \caption{\label{fig:wmconcept} 
    \textit{Four-Wave Mixing in a Quasi-1D Lattice.}
    (a) Qubit population from continuous-wave measurement exhibits three features: the $g-e$ transition at $\wq = 2\pi \times 9.15$~GHz, the $(g-f)/2$ (two-photon) transition at $2\pi \times 9.087$~GHz, and an additional spectral response at $\wps = 2\pi \times 9.211$~GHz corresponding to the four-wave-mixing process. The inset shows the lowest three transmon energy levels, illustrating the negative anharmonicity $-\alpha/2\pi = -125$~MHz.
    (b) Schematic of the four-wave-mixing process. Two input photons—the monitor photon at $\wm$ and the pump photon at $\wps$—combine to generate two outputs: a qubit photon $\wq$ and the undriven output photon at $\ww$. The diagram on the right shows all relevant frequencies, with coherent drives represented as delta functions and the other modes as Lorentzians. For the data shown in (a) with $\wm = 2\pi \times 9.697$~GHz, the output photon emerges at $\ww = 2\pi \times 9.758$~GHz.
    }
\end{figure}

\begin{figure*}[ht!]
\centering
    \includegraphics[width=0.98\textwidth]{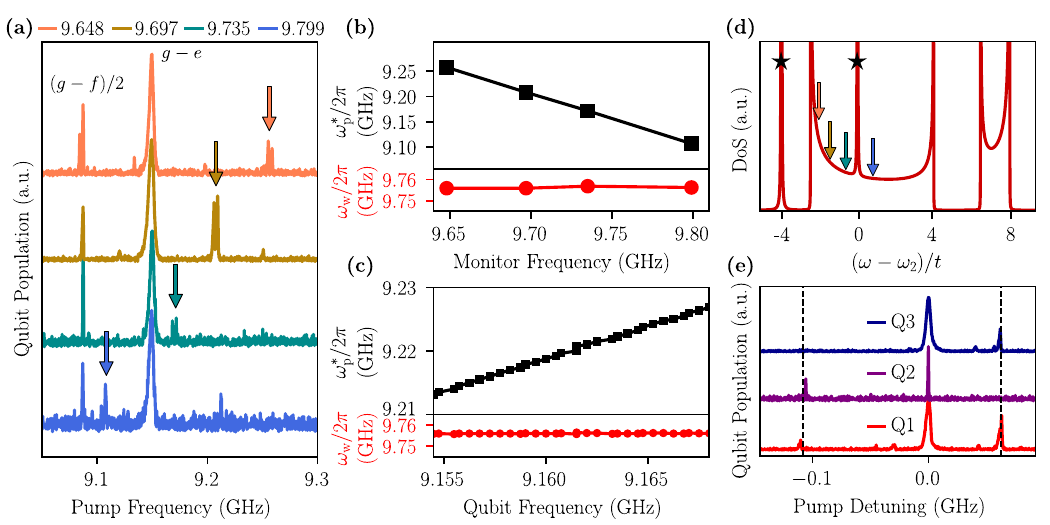}
    \vspace{-0.2cm}
    \caption{\label{fig:wmcharacterization} 
    \textit{Four-Wave-Mixing Characterization.}
    (a) Qubit population extracted for different monitor modes as a function of $\omega_p$. The monitor mode frequencies are indicated in the legend. The qubit frequencies $\wq = 2\pi \times 9.15$~GHz and the $(g-f)/2$ transition remain constant. The pump frequencies $\wps$ where the wave-mixing process is resonant are marked by colored arrows.
    (b)–(c) The resonant pump frequencies $\wps$ and calculated wave-mixing frequencies $\ww$ versus (b) $\wm$ and (c) $\wq$. In both cases, the output frequency calculated from Eq.~\eqref{eq:wm} remains constant at $\ww = 2\pi \times 9.758$~GHz.
    (d) Locations of the four monitor modes (arrows) and the two flat bands (stars) are indicated on the simulated density of states of the FW modes from Fig.~\ref{fig:FW_tuning}(b) in the tight-binding limit. The four monitor modes are color-coded to match corresponding data in (a).
    (e) The wave-mixing response for all three qubits and both flat bands. In addition to the $g-e$ transition, defined as a zero detuning, the upper flat band at $62$~MHz detuning---which contains only odd-parity modes---couples only to the off-axis qubits \sixteen and \eightns, but not to the on-axis qubit \fivens. In contrast, the lower flat band at $-108$~MHz detuning---which contains both even- and odd-parity modes---couples to the on-axis qubit \five and the off-axis qubit \sixteenns.
    }
\end{figure*}

\textit{Four-Wave-Mixing Process}.---In addition to the strong photon-photon coupling discussed above, superconducting circuits also exhibit strong qubit-photon coupling. In multimode systems, such as CPW lattices, this introduces new nonlinear processes that couple the qubit and localized photon modes, such as those found in the flat bands shown in Fig.~\ref{fig:FW_tuning}. A transmon qubit~\cite{Koch:transmon,Blais:CavityQED} typically exhibits a single-photon $g-e$ transition with frequency $\wq$ and a two-photon transition to the $\ket{f}$ state, denoted $(g-f)/2$, at a lower frequency $\wq - \alpha/2$, where $-\alpha$ is the anharmonicity. Here, we observe a qualitatively different nonlinear process beyond these two standard transitions, in which an undriven lattice mode and the qubit are excited via an interaction mediated by a weakly driven monitor mode and a strong pump tone, as shown in Fig.~\ref{fig:wmconcept}(b). We further show that, in our device, this process is dominated by contributions involving undriven modes belonging to the flat bands.

We use the conventional dispersive-readout method \cite{Blais:CavityQED, wallraff:dispersiveReadout, Blais:CircuitQED} that employs a pump-probe measurement scheme to observe the four-wave-mixing process. The measurement probes the transmission of a selected monitor mode at $\wm$ while the pump tone $\omega_\mathrm{p}$ is swept (see Appendix~\ref{app:wmappendixfull} for a detailed measurement scheme). Due to the presence of qubit-state-dependent frequency shifts, the measurement tone resonant with the monitor mode exhibits amplitude and phase shifts when the qubit is driven out of its ground state. Qubit excitation associated with the two standard transitions is observed at $\wq$ and $\wq - \alpha/2$, as shown in Fig.~\ref{fig:wmconcept}(a). In addition, we observe an extra qubit excitation appearing at a new frequency $\wps$, resulting from a four-photon scattering process, or equivalently, a four-wave-mixing process.

The four-wave-mixing process can be interpreted as a joint excitation of a qubit $\dqb$ and an undriven lattice mode $\fbop$, where $\fbop$ denotes the wave-mixing mode. This process is mediated by a weakly driven monitor mode $\mop$ and a strong pump tone with strength $\mathcal{E}_{\mathrm{p}}$. The governing Hamiltonian is
\begin{align}
H_0 = &H_\mathrm{res} + \wq \dqb^\dag \dqb 
- \frac{\alpha}{2} \dqb^\dag \dqb^\dag \dqb \dqb 
+ \left(\mathcal{E}_{\mathrm{p}} e^{-i \omega_\mathrm{p} t} \dqb^\dag
+ \mathcal{E}^*_{\mathrm{p}} e^{i \omega_\mathrm{p} t} \dqb \right) \nonumber
\\
& + \gqb (\dqb + \dqb^\dag)(\fbop + \fbop^\dag)
+ \gqa (\dqb + \dqb^\dag)(\mop + \mop^\dag),
\end{align}
where $H_\mathrm{res} = \ww \fbop ^\dag \fbop + \wm \mop ^\dag \mop$, and $\gqa$ and $\gqb$ denote the respective qubit–mode coupling strengths. The nonlinearity of the qubit induces effective coupling terms between the photon modes \cite{Devoret:firstwm, YvonneGao:thesis, Schoelkopf:earlywm}. In the observed four-wave-mixing process, shown schematically in Fig.~\ref{fig:wmconcept}(b), one monitor photon at $\wm$ and one pump photon at $\wps$ are converted into one qubit excitation at $\wq$ and one photon in the undriven output wave-mixing mode at $\ww$. In the rotating frame set by the pump frequency $\wps$, this process can be described by an effective interaction term $\mathcal{E} _{\mathrm{p}} \dqb ^\dag \fbop ^\dag \mop$ (see Appendix~\ref{app:wminttheory} for a full derivation). Accordingly, the process becomes resonant when the energy-conservation condition
\begin{equation}\label{eq:wm}
    \wm + \wps = \wq + \ww
\end{equation}
is satisfied. While the four-wave mixing can in principle occur between any two modes of the lattice, we find that only a subset of modes dominates the process, independent of the specific choice of monitor or qubit frequency. In our lattice, the flat bands host strongly coupled modes that exhibit the strongest four-wave-mixing resonances.

To verify that the observed behavior is consistent with Eq.~\eqref{eq:wm}, we characterize the four-wave-mixing resonance as a function of the monitor, qubit, and pump frequencies, as well as the saturation behavior of the qubit. When $\wq$ is tuned via an on-chip flux-bias line, the resonant pump frequency $\wps$ shifts by the same amount, as shown in Fig.~\ref{fig:wmcharacterization}(c). Further confirmation that only one pump photon is absorbed per qubit excitation can be found by examining the saturation behavior of the qubit as a function of pump power, shown in Appendix~\ref{app:wmappendixpowerscan}. The saturation behavior of the four-wave-mixing resonance is consistent with single-photon saturation, and is inconsistent with a two-photon process such as the $(g-f)/2$ transition. Unlike $\wq$, the monitor frequency can only be changed in discrete steps, by switching to a different lattice mode. The four monitor modes selected span a large portion of the FW band structure, and are indicated with colored arrows in Fig.~\ref{fig:wmcharacterization}(d). Spectroscopy data corresponding to these modes are shown in Fig.~\ref{fig:wmcharacterization}(a). In this case, $\wps$ exhibits a frequency shift equal and opposite to the change in the monitor frequency $\wm$, while the qubit transitions remain fixed, as shown in Fig.~\ref{fig:wmcharacterization}(a)-(b). These observations in Fig.~\ref{fig:wmcharacterization}(a)-(c) are consistent with Eq.~\eqref{eq:wm} and output a single fixed frequency $\ww = 2\pi \times 9.758$~GHz.

The microscopic model of the four-wave-mixing process, presented in Appendix~\ref{app:wminttheory}, shows that the strength of the process scales linearly with $\gqb$ of the undriven cavity mode. Therefore, the first resonances that emerge are expected to correspond to the most strongly-coupled modes. In most discrete translation-invariant lattice systems, all modes are delocalized. However, CPW lattices naturally give rise to flat bands~\cite{Kollar:2019linegraph, Kollar:2019hyperbolic}, supporting localized photonic modes with very strong coupling to qubits. The output frequency $\ww = 2\pi \times 9.758$~GHz found in Fig.~\ref{fig:wmcharacterization}(a)-(c) corresponds to the predicted location of one of the flat bands in the lattice~\cite{latticepaper}. Below, we use the reflection symmetry of the lattice and qubits at multiple locations to confirm that the strongly-coupled modes originating from the flat bands dominate the four-wave-mixing process. 

The FW modes of the quasi-1D lattice contain two distinct sets of flat bands, as shown in Fig.~\ref{fig:FW_tuning}. For these modes, both the previous tight-binding simulations~\cite{latticepaper} and the circuit theory presented above, obey a reflection symmetry about the horizontal axis (see Appendix~\ref{app:quasi-1Dlattice}). The upper flat band contains only odd-parity modes with no amplitude on the reflection axis. As a result, these modes couple only to qubits located on off-axis sites. In contrast, the lower flat band contains both even- and odd-parity modes with significant amplitude on the on-axis resonators. Thus, the lower flat band couples strongly to the on-axis qubit, in addition to the off-axis qubits. Four-wave-mixing resonances for all three qubits, with output at both flat bands, are shown in Fig.~\ref{fig:wmcharacterization}(e). The higher-frequency resonance ($\ww = 2\pi \times 9.758$~GHz), corresponding to output at the upper flat band, appears only for the off-axis qubits \sixteen and \eightns. In contrast, the lower-frequency resonance ($\ww = 2\pi \times 9.589$~GHz), corresponding to output at the lower flat band, is observed both for the on-axis qubit \five and off-axis qubit \sixteenns.

\vskip 0.1in

\textit{Discussion}.---The multimode strong-coupling effects presented here were derived and observed for the specific case of transmon qubits in a microwave resonator lattice; however, the observed four-wave-mixing process and beyond-tight-binding coupling terms readily generalize to any spin–boson system, including other types of superconducting qubits, as well as atomic cavity QED systems and photonic platforms. The four-wave-mixing process provides a new spectroscopic pathway for measuring localized modes in multimode systems, and could be used to test Anderson localization in hyperbolic lattices \cite{Kollar:2019hyperbolic, Kollar:2019linegraph, Boettcher:anderson, Victor:weaklocalization}. Conversely, the output mode of the wave-mixing process can be selected by choosing pump and monitor frequencies that are frequency matched to the desired output. This enables the production of antibunched nonclassical light \cite{Collins:singlephoton, steckquantum} in the selected mode, independent of the qubit (two-level system) frequency, and the production of on-chip entangled photon pairs~\cite{Thompson:entanglement} with wavelength multiplexing. While derived specifically for the case of a superconducting circuit Lagrangian, the qualitative alterations to the band structure observed here arise from two phenomena which are general properties of coupled atoms/oscillators: Rabi-model-like coupling terms $(a^\dagger + a)(b^\dagger + b) \neq a^\dagger b + b^\dagger a$ and mixing of on-site wavefunctions due to the perturbation of the coupler. The correction terms derived here can profoundly alter strongly coupled microwave resonator arrays~\cite{Bosman:ultrastrongcoupling, vrajitoarea2024, Painter_Markov, Painter_SSH} and can control the degree of degeneracy achievable in slow-light photonics~\cite{chang:RevModPhys}, where flat (or nearly flat) bands are used to enhance light-matter interactions.

\vskip 0.1in

\begin{acknowledgements}
This research was supported by the Air Force Office of Scientific Research (Grant No. FA9550-21-1-0129), the National Science Foundation (QLCI grant OMA-2120757, PHY2047732, and PFC at JQI PHY-1430094), the Sloan Foundation, and the Maryland Quantum Materials Center. A.F., A.B.,  and A.V.G.~were supported in part by ARL (W911NF-24-2-0107), NSF QLCI (award No.~OMA-2120757), the DoE ASCR Quantum Testbed Pathfinder program (award No.~DE-SC0024220), ONR MURI,  NSF STAQ program, AFOSR MURI,  DARPA SAVaNT ADVENT,  and NQVL:QSTD:Pilot:FTL. A.F., A.B.,  and A.V.G.~also acknowledge support from the U.S.~Department of Energy, Office of Science, National Quantum Information Science Research Centers, Quantum Systems Accelerator (award No.~DE-SCL0000121) and from the U.S.~Department of Energy, Office of Science, Accelerated Research in Quantum Computing, Fundamental Algorithmic Research toward Quantum Utility (FAR-Qu). Y.-X.W.~acknowledges support from a QuICS Hartree Postdoctoral Fellowship. 
\end{acknowledgements}

\bibliographystyle{apsrev4-2}
\bibliography{refs.bib}






\clearpage
\onecolumngrid

\makeatletter
\let\addcontentsline\latex@addcontentsline
\makeatother

\setcounter{figure}{0} 
\setcounter{equation}{0}
\setcounter{section}{0}
\setcounter{tocdepth}{2}

\renewcommand{\figurename}{Figure} 
\renewcommand{\thefigure}{S\arabic{figure}} 
\renewcommand{\thetable}{S\arabic{table}} 
\renewcommand{\theequation}{S\arabic{equation}} 
\renewcommand{\thesection}{S\arabic{section}}

\renewcommand{\theHfigure}{S\arabic{figure}}
\renewcommand{\theHequation}{S\arabic{equation}}
\renewcommand{\theHsection}{S\arabic{section}}

\begin{center}
    \Large Supplemental Material:\\
    Multimode Strong-Coupling Processes in Circuit QED Lattices
\end{center}

\tableofcontents

\vskip 0.6in

\section{Quasi-1D CPW Lattice Device}\label{app:quasi-1Dlattice}

\begin{figure}[h]
\centering
    \includegraphics[width=\textwidth]{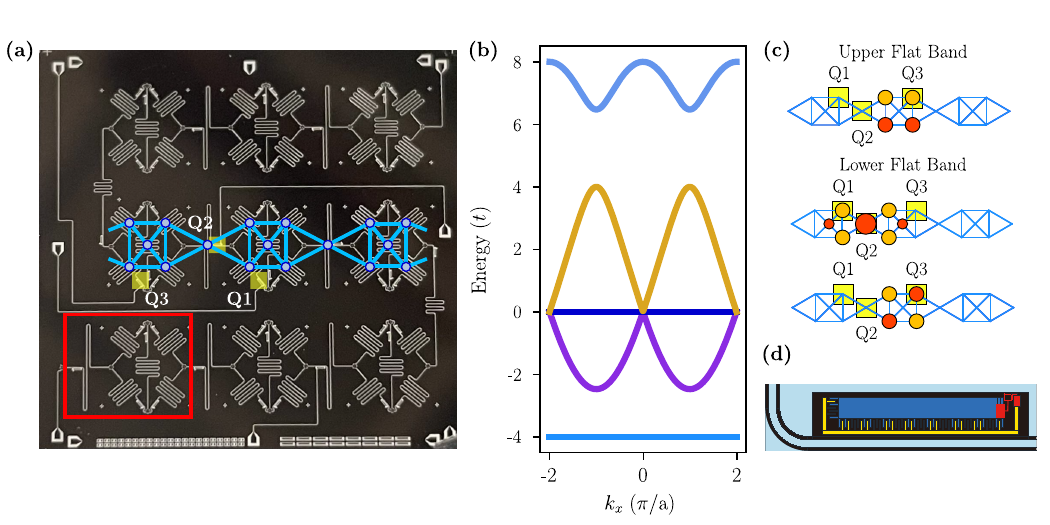}
    \vspace{-0.2cm}
    \caption{\label{fig:device} 
    \textit{Device Design and Band Structure.}
    (a) Photograph of quasi-1D CPW lattice device. CPW resonators shown in white. The device consists of nine identical unit cells; one unit cell is highlighted in red. Three transmon qubits are indicated by yellow boxes with labels. The effective connections between resonator sites of the middle row are highlighted by blue lines.
    (b) The band structure of full-wave modes from the tight-binding simulation expressed in units of the hopping strength $t$. The energy of an uncoupled resonator is set to zero. The band structure contains three distinct sets of bands: a highest isolated dispersive band, a middle set with an upper flat band and two dispersive bands, and two doubly-degenerate flat bands at the bottom.
    (c) Schematic of representative normal modes of the two flat bands in the full-wave regime, with only the central three unit cells containing the qubits (highlighted by yellow boxes) shown. The size and color of each circle represent the amplitude and phase of the mode function, respectively. On-axis resonators are located along the reflection axis, and off-axis resonators are peripheral resonators within each unit cell. The full-wave modes exhibit two types of flat bands with different reflection symmetry. The upper flat band shows odd parity and is strongly localized on the off-axis resonators. In contrast, the degenerate lower flat bands contain both odd and even parity modes, which together have strong overlap with all sites in the unit cell.
    (d) A flux-tunable transmon is capacitively coupled to a CPW resonator. Every resonator contains capacitor paddles (shown in blue and yellow); if it is an active qubit, a DC superconducting quantum interference device (DC-SQUID, shown in red) is also included.
    }  
\end{figure}

In this section, we describe in detail the quasi-1D CPW lattice device studied in this work. For further information, particularly regarding device characterization and analysis, we refer the reader to Ref.~\cite{latticepaper}. The device exploits the flexible connectivity of CPW resonators to realize a wide variety of bands, while incorporating multiple qubits to enable photon-mediated qubit–qubit interactions. 

The device contains nine rhombus-shaped unit cells with six CPW resonators each, which are arranged in a one-dimensional chain, as shown in Fig.~\ref{fig:device}(a). Three transmon qubits, labeled \sixteenns, \fivens, and \eightns, are coupled to individual CPW resonators in the lattice. The locations of these transmons in the device are highlighted with yellow boxes in Fig.~\ref{fig:device}(a), and the design of the transmon is shown in Fig.~\ref{fig:device}(d). These transmon qubits are tuned via on-chip flux-bias lines from the half-wave mode frequency (approximately $4.8$~GHz) to the full-wave mode frequency (approximately $9.6$~GHz), covering the entire frequency range of interest. The resonators, all with equal resonant frequency, are connected through identical three-way couplers to ensure uniform nearest-neighbor hopping strength, aside from small fabrication-induced disorder. Measurements are performed by probing transmission of microwave signal through the device, with input/output ports located at the lower-left and upper-right corners of the device, shown in Fig.~\ref{fig:device}(a). Appendix~\ref{app:wmappendixfull} details the measurement protocol.

Due to the connectivity described above, both the half-wave and full-wave modes of the device exhibit nontrivial band structures. The band structure of the lattice can be obtained from a tight-binding model \cite{Kollar:2019linegraph, Koch:AnnPhysBerl, Koch:TRS_breaking, Houck_Nature_QS, latticepaper}, which is valid only for small hopping strengths, or, in general, from the full nonlinear multimode circuit analysis discussed in the main text and Appendix~\ref{app:bandcalculation}. The full-wave band structure in the tight-binding limit, shown in Fig.~\ref{fig:device}(b), contains an isolated dispersive band at the top, two dispersive bands with a linear crossing involving the upper flat band, and doubly-degenerate gapped flat bands at the bottom. Such gapped flat bands, which are not present in generic lattices, naturally arise in CPW lattices with non-bipartite connectivity \cite{Kollar:2019hyperbolic, Kollar:2019linegraph}. The band structure of the half-wave modes is presented in Appendix~\ref{app:bandcalculation}. Higher CPW harmonics give rise to band structures similar to those of the half-wave and full-wave modes, differing primarily in the hopping strength. The operating frequencies of the tunable transmon qubits lie well below these higher bands, so they will not be considered further, except with regard to the circuit Lagrangian calculation in Appendix~\ref{app:bandcalculation}.

The full-wave modes exhibit two types of flat bands, distinguished by a reflection symmetry about the horizontal axis. The spatial mode functions for representative modes in these flat bands are shown in Fig.~\ref{fig:device}(c). In the absence of resonator disorder, the upper flat-band consists of a single odd-parity band, with mode functions highly localized on off-axis resonators (i.e., peripheral resonators within the unit cell). The lower flat bands are doubly degenerate, with one even-parity band and one odd-parity band. The even-parity modes overlap with the on-axis resonators.


The qubit-location dependence of the four-wave-mixing resonances discussed in the main text is consistent with the reflection symmetry of the flat bands predicted by the tight-binding model. Four-wave-mixing processes can occur only when there is coupling between the qubit and the output mode, i.e., when the lattice mode function is nonzero at the qubit site. The mode functions of the upper and lower sets of flat bands overlap with different subsets of the lattice cites, and different subsets of \sixteenns, \fivens, \eight [see  Fig.~\ref{fig:device}(c)]. As seen in Fig.~\ref{fig:wmcharacterization}(e), the observed four-wave-mixing resonances of the upper flat band are associated with the off-axis qubits (\sixteen and \eightns), consistent with the predicted odd parity of this band. In contrast, the four-wave-mixing resonances of the lower flat band are associated with both on-axis (\fivens) and off-axis qubits (\sixteenns), consistent with the presence of an even-parity band.


In this device, the flat-band states are highly localized and have very little overlap with the measurement ports. As a result they do not produce strong signatures in transmission, as was observed in Ref.~\cite{latticepaper}.  However, the four-wave-mixing process observed here probes these modes without the need for significant direct input-output coupling; see Fig.~\ref{fig:wmconcept} and Fig.~\ref{fig:wmcharacterization}.

\section{Band Calculation for the Lattice Device}\label{app:bandcalculation}

This section describes the theoretical modeling of the nonlinear multimode circuit-QED system discussed in the main text. Our system consists of a multimode coplanar-waveguide (CPW) resonator lattice coupled to transmon qubits. Previous works have treated related systems as linear tight-binding resonator lattices coupled to individual qubits, in which counter-rotating wave-mixing interactions within the lattice and higher-order dispersive effects induced by the qubit anharmonicities are typically ignored. In our system, however, the bandwidth of the CPW lattice can be comparable to the fundamental mode frequency of the resonators, so such higher-order nonlinearities become essential for understanding the system's dynamics, especially the four-wave-mixing processes studied in the main text. We model the system using the lumped-element circuit approach. In what follows, we first derive the band structure and normal modes of the resonator lattice. We then calculate the coupling between the resonator-lattice normal modes and the qubits.

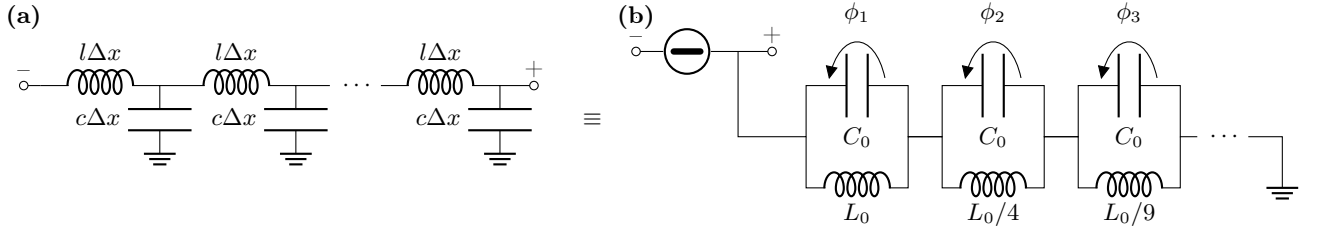
\begin{figure}[t]
    \centering
    \begin{circuitikz}[scale=0.9]

        \draw (0, 1) node[]{$\textbf{(a)}$} to (0, 1);
        
        \draw (0, 0) node[anchor=south]{--} to [short, o-] (0.25, 0);
        \draw (0.25, 0) to [L, l^=$l\Delta x$] (2, 0) to [C, l_=$c\Delta x$] (2, -1) node[tlground]{};
        \draw (2, 0) to (2.25, 0) to [L, l^=$l\Delta x$] (4, 0) to [C, l_=$c\Delta x$] (4, -1) node[tlground]{};
        \draw (4, 0) to (4.5, 0);
        \draw[white] (4.6, 0) to [short, l={$\cdots$}, black, label distance=-9pt](5.25, 0);
        \draw (5.25, 0) to [L, l^=$l\Delta x$] (7, 0) to [C, l_=$c\Delta x$] (7, -1) node[tlground]{};
        \draw (7, 0) to [short, -o] (7.5, 0) node[anchor=south]{+};

        \draw (8,0)[white] to [short, l={$\equiv$}, black](8, -1);

        \draw (9, 1) node[]{$\textbf{(b)}$} to (9, 1);

        \draw (9, 0.5) node[anchor=south]{--} to [short, o-](9.4, 0.5);
        \draw (9.75, 0.5) to (9.75, 0.5)node[vsourceCshape, scale=0.7, black]{};
        \draw (10.1, 0.5) to [short, -o](11, 0.5)node[anchor=south]{+};
        \draw (10.5, 0.5) to (10.5, -0.75) to (11.5, -0.75) to (11.5, 0)
        to [C, l_=$C_0$, v^<=$\phi_1$](13, 0) to (13, -0.75) to (13.5, -0.75) to (13.5, 0)
        to [C, l_=$C_0$, v^<=$\phi_2$](15, 0) to (15, -0.75) to (15.5, -0.75) to (15.5, 0)
        to [C, l_=$C_0$, v^<=$\phi_3$](17, 0) to (17, -0.75) to (17.25, -0.75);
        \draw[white] (17.35, -0.75) to [short, l={$\cdots$}, black, label distance=-9pt](18, -0.75);
        \draw (18, -0.75) to (18.5, -0.75) to (18.5, -1.5)node[tlground]{};
        \draw (11.5, -0.75) to (11.5, -1.5)
        to [L, l_=$L_0$](13, -1.5) to (13, -0.75) to (13.5, -0.75) to (13.5, -1.5)
        to [L, l_=$L_0/4$](15, -1.5) to (15, -0.75) to (15.5, -0.75) to (15.5, -1.5)
        to [L, l_=$L_0/9$](17, -1.5) to (17, -0.75);
        
    \end{circuitikz}
    \caption{
    \textit{Circuit Diagrams of a Coplanar-Waveguide Resonator.}
    (a) A coplanar-waveguide resonator can be modeled as a series of inductors that are capacitively coupled to the ground.
    (b) The normal modes of this circuit are standing waves along it that correspond to a series of LC resonators.}
    \label{fig:resonator_circuit}
\end{figure}

A key building block of our device is a CPW resonator. A single CPW resonator can be modeled as a series of small inductors capacitively coupled to  ground along the length of the resonator [see Fig.~\ref{fig:resonator_circuit}(a), where $l$ and $c$ are the inductance and capacitance per unit length of the resonator, respectively]. This circuit exhibits standing waves equivalent to those of a circuit composed of a series of LC resonators with $C_j=C_0$ and $L_m=L_0/j^2$ for the $j^{\text{th}}$ LC resonator in the series [Fig.~\ref{fig:resonator_circuit}(b)]. The $j^{\text{th}}$ LC resonator has angular frequency $\omega_j=\frac1{\sqrt{L_jC_j}}=j\omega_0$ where $\omega_0=\frac1{\sqrt{L_0 C_0}}$. The fundamental normal-mode frequency is related to the finite-element description in Fig.~\ref{fig:resonator_circuit}(a) via $\omega_0=\frac{\pi}{L\sqrt{lc}}$, where $L$ is the length of the resonator. We are specifically interested in the first two harmonics: the half-wave modes ($j=1$) and the full-wave modes ($j=2$).

The corresponding Lagrangian for the circuit in Fig.~\ref{fig:resonator_circuit}(b) is
\begin{equation} \label{eq:res_Lagr}
	\mathcal L_{\rm res} = \sum_{j=1}^\infty \left( \frac{C_0}{2} \dot \phi_{j}^2 - \frac{j^2}{2L_0} \phi_{j}^2 \right).
\end{equation}
Here, $\phi_j$ is the flux through the $j^{\text{th}}$ LC resonator. The total fluxes at the $+$ and $-$ ends of the resonator are $\phi^+ = \sum_j \phi_{j}$ and $\phi^- = \sum_j (-1)^j \phi_{j}$, respectively. The choice of which end is labeled $+$ or $-$ is arbitrary, since the minus signs can be absorbed into the flux definitions. The \begin{circuitikz}[baseline=-0.6ex] \draw (0,0) to (0,0) node[vsourceCshape, scale=0.5]{};\end{circuitikz} component represents the odd-$j$ modes (i.e., modes with half-integer wavelengths along the resonator) having opposite fluxes at the two ends.

\begin{figure}[h]
    \centering
    \begin{circuitikz}[scale=0.9]
        \draw[white] (-.75, 0) to [short, l={$\cdots$}, black, label distance=-9pt](0, 0);
        \draw (0, 0) node[anchor=south]{$\overline{P_i}$} to [european resistor, o-o, l={${\rm CPW}_i$}, label distance=-12pt] (2, 0) node[anchor=south]{$P_i$};
        \draw (2, 0) to [C, l_=$C_c$, o-o] (3, 0);
        \draw (3, 0) node[anchor=south]{$P_{i'}$} to [european resistor, o-o, l={${\rm CPW}_{i'}$}, label distance=-12pt] (5, 0) node[anchor=south]{$\overline{P_{i'}}$};
        \draw[white] (5, 0) to [short, l={$\cdots$}, black, label distance=-9pt](5.75, 0);
    \end{circuitikz}
    \caption{\textit{Connection of Two CPW resonators $i$ and ${i'}$ by a Coupling Capacitor $C_c$.} $P_i, P_{i'} \in \{+, -\}$ denote the parities of the resonator ends connected by the capacitor.}
    \label{fig:coupled_resonators}
\end{figure}
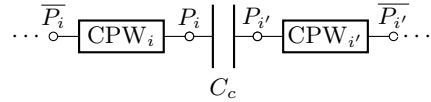

Two CPW resonators can be coupled capacitively at their ends, as shown in Fig.~\ref{fig:coupled_resonators}. The coupled ends of the resonators can have either $+$ or $-$ parity, denoted by $P_i$ and $P_{i'}$. The other end of each resonator has the opposite parity; i.e., $\overline{P_{i/{i'}}} = -P_{i/{i'}}$.

The Lagrangian for the circuit shown in Fig.~\ref{fig:coupled_resonators} is
\begin{equation}\label{eq:res_coupling}
	\mathcal L_{\rm c.res.}(i, {i'}) = \mathcal L_{\rm res}(i) + \mathcal L_{\rm res}({i'}) + \frac{C_c}{2} \left( \dot \phi_i^{(P_i)} - \dot \phi_{i'}^{(P_{i'})} \right)^2,
\end{equation}
where $\mathcal L_{\rm res}(i)$ is the single-resonator Lagrangian from Eq.~\eqref{eq:res_Lagr} with $\phi_{i, j}$ substituted for $\phi_j$. 

\begin{figure}[b!]
    \centering
    \begin{circuitikz}[scale=0.9]
        \draw[white] (-.75, 0) to [short, l={$\cdots$}, black, label distance=-9pt](0, 0);
        \draw (0, 0) to (0.5, 0) to (0.5, 1) to (0.7, 1) to (0.7, 0.4) to [barrier, l=$E_{J0}$, label distance=-25pt, bipoles/length=1.2cm](2.3, 0.4) to (2.3, 1) to (2.5, 1) to (2.5, 0) to (3, 0) to (3, -0.5)node[tlground]{};
        \draw (0.7, 1) to (0.7, 1.6) to [barrier, l^=$E_{J0}$, label distance=-3mm, bipoles/length=1.2cm, v^<=$\chi + \Phi_\text{ext}/2$](2.3, 1.6) to (2.3, 1);
        \draw[white] (0.7, 1) to [short, l=$\Phi_{\text ext}$, black, label distance=-9pt](2.3, 1);
        \draw (0.5, 0) to (0.5, -1) to [C, l_=$C_q$](2.5, -1) to (2.5, 0);

    \end{circuitikz}
    \caption{
    \textit{Circuit Representation of a Transmon Qubit.}
    A transmon qubit has an effective flux $\chi$ through the DC-SQUID. The external magnetic flux $\Phi_\text{ext}$ tunes the effective Josephson energy $E_{J} = 2E_{J0} \cos \left( \pi {\Phi_\text{ext}}/{\phi_0} \right)$.
    }
    \label{fig:transmon_circuit}
\end{figure}
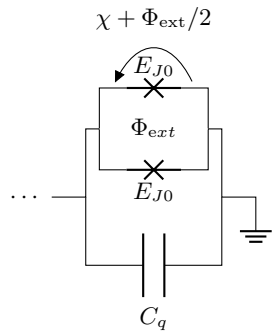

Three resonators in our device are also coupled to transmon qubits at one end. A flux-tunable transmon consists of a DC superconducting quantum interference device (DC-SQUID) shunted by a capacitor, shown in Figs.~\ref{fig:device}(d) and \ref{fig:transmon_circuit}. To reduce disorder, every resonator includes the capacitor paddles that could host a transmon, but only three of them include a DC-SQUID. However, we can assume all of them include a DC-SQUID and are functional transmons that are far detuned from the half-wave and full-wave modes unless one of them is detuned closer to these modes. This helps with treating every unit cell in the lattice uniformly.

Each transmon can be described by the following Lagrangian:
\begin{equation}
    \mathcal L_{\rm tm}(q) = \frac{C_q}{2} \dot \chi^2 + 2 E_{J0} \cos \left( \pi \frac{\Phi_\text{ext}}{\phi_0} \right) \cos\left( 2\pi \frac{\chi}{\phi_0} \right)
    = \frac{C_q}{2} \dot \chi^2 + E_{J} \cos\left( 2\pi \frac{\chi}{\phi_0} \right),
\end{equation}
where $\phi_0 = \frac{2\pi}{2e}$ denotes the flux quantum for a Cooper pair (note that we have set $\hbar=1$), $\chi \pm \Phi_\text{ext}/2$ is the flux through the Josephson junctions resulting in a flux of $\Phi_{\text{ext}}$ through the DC-SQUID loop, and $E_{J} = 2 E_{J0} \cos \left( \pi {\Phi_\text{ext}}/{\phi_0} \right)$ is the effective Josephson energy of the DC-SQUID (see Fig.~\ref{fig:transmon_circuit}). In what follows, we treat the transmon as having a single junction with Josephson energy $E_{J}$. Expanding the cosine function $\cos\left( 2\pi  {\chi}/{\phi_0} \right)$ around zero flux (and dropping the constant term) yields the approximate Lagrangian up to $O(\chi^4)$:
\begin{equation} \label{eq:tranmon}
	\mathcal L_{\rm tm}(q) \simeq \frac{C_q}{2} \dot \chi^2 - \frac{1}{2L_q} \chi^2 + \tilde \alpha \chi^4,
\end{equation}
with 
\begin{equation}\label{eq:transmon_params}
    L_q = \left( \frac{1}{2e} \right)^2 \frac{1}{E_J}, \tilde \alpha = \left( 2e \right)^4 \frac{E_J}{24}.
\end{equation}

The Lagrangian in Eq.~\eqref{eq:tranmon}, up to neglecting counter-rotating terms, corresponds to an anharmonic oscillator with frequency $\omega_q = 1/\sqrt{C_q L_q}$ and anharmonicity $\alpha$ that is proportional to $\tilde \alpha$.

Each transmon qubit $q_i$ is coupled to its adjacent resonator $i$ via a capacitor $C_c'$, resulting in the coupling contribution to the Lagrangian:
\begin{equation}\label{eq:res_q_coupling}
	\mathcal L_{\rm res-tm.cp.}(i, q_i) = \frac{C_c'}{2} \left( \dot \phi_i^{(P_i)} - \dot \chi_i \right)^2.
\end{equation}
Here, $P_i$ is the parity of the end of the resonator that the transmon is coupled to.

Having introduced all the elements of our CPW-transmon lattice and assuming  periodic boundary conditions, we can write down the total Lagrangian of the device:
\begin{equation}\label{eq:lattice_lag}
	\mathcal L_{\rm lattice} = \sum_{i=1}^N \mathcal L_{\rm res}'(i) - C_c \sum_{\left< i, {i'} \right>} \dot \phi_i ^{(P_i^{i'})} \dot \phi_{i'} ^{(P_{i'}^i)} + \sum_{i=1}^N \mathcal L_{\rm tm}'(q_i) - C_c' \sum_{i=1}^N \dot \phi_i^+ \dot \chi_i.
\end{equation}
Here, we have introduced the notation $P_i^{i'}$ to denote the parity of resonator $i$ at the end coupled to neighboring resonator ${i'}$. The end parities of each resonator are defined such that the transmon is coupled to the $+$ end. We have redefined the single-resonator and single-transmon Lagrangians by absorbing the single-flux coupling terms:
\begin{equation}
	\mathcal L_{\rm res}'(i) = \sum_{j=1}^\infty \left( \frac{C_0}{2} \dot \phi_{i, j}^2 - \frac{j^2}{2L_0} \phi_{i, j}^2 \right) + C_c \left( (\dot \phi_i^+)^2 + (\dot \phi_i^-)^2 \right) + \frac{C_c'}{2} \left( \dot \phi_i^+ \right)^2,
\end{equation}
and
\begin{equation}
	\mathcal L_{\rm tm}'(q_i) = \frac{C_q+C_c'}{2} \dot \chi_i^2 - \frac{1}{2L_q} \chi_i^2 + \tilde \alpha \chi_i^4.
\end{equation}
In our system, each bulk resonator is coupled to two other resonators at its $+$ end and two other resonators at its $-$ end [see Fig.~\ref{fig:device}(a) and also the inset in Fig.~\ref{fig:FW_tuning}(a) for the layout and resonator connectivity of the device]. 
Therefore, with periodic boundary conditions, $\mathcal L_{\rm res}'$ will have an identical form for all resonators.

We are interested in the low-energy behavior of Eq.~\eqref{eq:lattice_lag}. Specifically, we focus on the half-wave ($j=1$) and full-wave ($j=2$) modes and transmons operating close to these modes. To that end, we break the calculation into the following steps: 

(i) approximately solving the normal modes of the resonator lattice after truncating the onsite modes to include only the first $M$ harmonics,

(ii) diagonalizing each transmon Lagrangian separately, and

(iii) treating the resonator-transmon couplings perturbatively.

The high-energy truncation in (i) should be chosen such that the low-energy modes have converged. The resonator-lattice part is quadratic and can be rewritten as
\begin{equation}\label{eq:res_lattice}
	\mathcal L_{\rm res-lattice} = \sum_{i=1}^N \mathcal L_{\rm res}'(i) - C_c \sum_{\left< i, {i'} \right>} \dot \phi_i ^{(P_i^{i'})} \dot \phi_j ^{(P_{i'}^i)} \equiv \frac{1}{2} \dot \Phi^{\rm T} \mathbf C \dot \Phi - \frac{1}{2} \Phi^{\rm T} \mathbf L \Phi,
\end{equation}
where $\Phi^{\rm T} = (\phi_{1, 1}, \phi_{1, 2}, \dots, \phi_{1, M}, \phi_{2, 1}, \dots, \phi_{N, M})$, and $\mathbf C$ and $\mathbf L$ are $NM\times NM$ matrices.

Noting that 
\begin{equation}
    \mathbf L = \mathbf I_N \otimes {\rm diag}\left( \frac{1^2}{L_0}, \frac{2^2}{L_0}, \dots, \frac{j^2}{L_0}, \dots, \frac{M^2}{L_0} \right)
\end{equation}
(with $\mathbf I_N$ the $N\times N$ identity matrix) is diagonal, we can find the normal modes of Eq.~\eqref{eq:res_lattice} by first rescaling the vector $\Phi$ to transform 
the coefficient matrix of $\mathbf{L}$ into an identity matrix, and then diagonalizing the remaining $\dot \Phi^{\rm T} \mathbf C \dot \Phi$ term. Thus, we define
\begin{equation}
	\tilde \Phi = \sqrt{\mathbf L} \Phi.
\end{equation}
The Lagrangian in Eq.~\eqref{eq:res_lattice} can then be rewritten in terms of $\tilde \Phi$ as
\begin{equation}
	\mathcal L_{\rm res-lattice} = \frac{1}{2} \dot{\tilde \Phi}^{\rm T} \frac{1}{\sqrt{\mathbf L}} \mathbf C \frac{1}{\sqrt{\mathbf L}} \dot{\tilde \Phi} - \frac{1}{2} \tilde \Phi^{\rm T} \mathbf I_{NM} \tilde \Phi,
\end{equation}
where $\mathbf I_{NM}$ is the $NM\times NM$ identity matrix. Since the second term now contains only the identity matrix, we can rotate $\tilde \Phi$ so that it diagonalizes $\mathbf K = \frac{1}{\sqrt{\mathbf L}} \mathbf C \frac{1}{\sqrt{\mathbf L}}$ in the first term:
\begin{equation}
	\mathcal L_{\rm res-lattice} = \frac{1}{2} \dot \Theta^{\rm T} \mathbf D \dot \Theta - \frac{1}{2} \Theta^{\rm T} \mathbf I_{NM} \Theta.
\end{equation}
Here, $\mathbf D = \mathbf V \mathbf K \mathbf V^{\rm T} = {\rm diag}(d_{1, 1}, d_{1, 2}, \dots, d_{1, M}, d_{2, 1}, \dots, d_{N, M})$ is a diagonal matrix and $\Theta = \mathbf V \tilde \Phi$. Finally,
\begin{equation}\label{eq:res_lattice_diagonal}
	\mathcal L_{\rm res-lattice} = \frac{1}{2} \sum_{i=1, j=1}^{N, M} \left( d_{i, j} \dot \theta_{i, j}^2 - \theta_{i, j}^2 \right).
\end{equation}
Here, the $\theta_{i, j}$ are the elements of $\Theta$. There are no longer mode-mode couplings between $\theta_{i, j}$'s, and each corresponds to a resonator-lattice normal mode with a frequency $\omega_{i, j} = \frac{1}{\sqrt{d_{i, j}}}$.

The transformation from $\phi_{i, j}$ to $\theta_{i, j}$ gives the following expression for $\phi_i^+$, the flux at the end of resonator $i$ coupled to the transmon qubit:
\begin{equation}
	\phi_i^+ = \sum_{j=1}^M \phi_{i, j} = \sum_{j=1}^M \sum_{i'=1, j'=1}^{N, M} \left( \frac{1}{\sqrt{\mathbf L}} \mathbf V^{\rm T} \right)_{(i, j), (i', j')} \theta_{i', j'},
\end{equation}
which lets us rewrite Eq.~\eqref{eq:lattice_lag} as
\begin{nalign}
	\mathcal L_{\rm lattice} = \frac{1}{2} \sum_{i=1, j=1}^{N, M} \left( d_{i, j} \dot \theta_{i, j}^2 - \theta_{i, j}^2 \right) + \sum_{i = 1}^N \mathcal L'_{\rm tm}(q_i) - C_c' \sum_{i,i'=1, j,j'=1}^{N, M} \left( \frac{1}{\sqrt{\mathbf L}} \mathbf V^{\rm T} \right)_{(i, j), (i', j')} \dot \theta_{i', j'} \dot \chi_i.
\end{nalign}
The corresponding Hamiltonian is obtained by a Legendre transformation:
\begin{nalign}\label{eq:lattice_Ham}
	\mathcal H_{\rm lattice} = & \sum_{i, j} \dot \theta_{i, j} \frac{\partial \mathcal L_{\rm lattice}}{\partial \dot \theta_{i, j}} + \sum_{i} \dot \chi_i \frac{\partial \mathcal L_{\rm lattice}}{\partial \dot \chi_i} - \mathcal L_{\rm lattice} \\
	= & \mathcal H_{\rm res-lattice} + \sum_{i = 1}^N \mathcal H_{\rm tm}(q_i) - C_c' \sum_{i,i', j,j'} \left( \frac{1}{\sqrt{\mathbf L}} \mathbf V^{\rm T} \right)_{(i, j), (i', j')} \dot \theta_{i', j'} \dot \chi_i,
\end{nalign}
where
\begin{equation}\label{eq:res_lattice_Ham}
	\mathcal H_{\rm res-lattice} = \frac{1}{2} \sum_{i=1, j=1}^{N, M} \left( d_{i, j} \dot \theta_{i, j}^2 + \theta_{i, j}^2 \right) = \sum_{i=1, j=1}^{N, M} \omega_{i, j} \left( b_{i, j}^{\dagger}b_{i, j}+\frac{1}{2} \right)
\end{equation}
and
\begin{equation}
	\mathcal H_{\rm tm}(q_i) = \frac{C_q+C_c'}{2} \dot \chi_i^2 + \frac{1}{2L_q} \chi_i^2 - \tilde \alpha \chi_i^4 = \frac{1}{2} \frac{n_i^2}{C_q + C_c'} + \frac{1}{2L_q} \chi_i^2 - \tilde \alpha \chi_i^4.
\end{equation}
Here, the $b_{i, j}^{\dagger}$ are the creation operators associated with the modes $\theta_{i, j}$, and $n_i = \partial \mathcal L_{\rm lattice} / \partial \dot \chi_i = {(C_q+C_c') \dot \chi_i}$ is the variable conjugate to $\chi_i$ and corresponds to the number of Cooper pairs of the transmon. The transmon Hamiltonian can be readily diagonalized to obtain energy levels of an isolated transmon:
\begin{equation}
	\mathcal H_{\rm transmon}(q_i) = \Omega_i \left( a_i^{\dagger}a_i + \frac{1}{2} \right) - \frac{\alpha}{2} a_i^{\dagger} a_i^{\dagger} a_i a_i.
\end{equation}
The transmon frequency $\Omega_i = \Omega_i(C_q+C_c', L_q, \tilde \alpha)$ is tunable using a flux line that tunes $E_J$ [and thus $L_q$ and $\tilde \alpha$, see Eq.~\eqref{eq:transmon_params}] and the transmon anharmonicity $\alpha \propto \tilde \alpha$.

\begin{figure*}[t]
    \centering
    \includegraphics[width=0.97\linewidth]{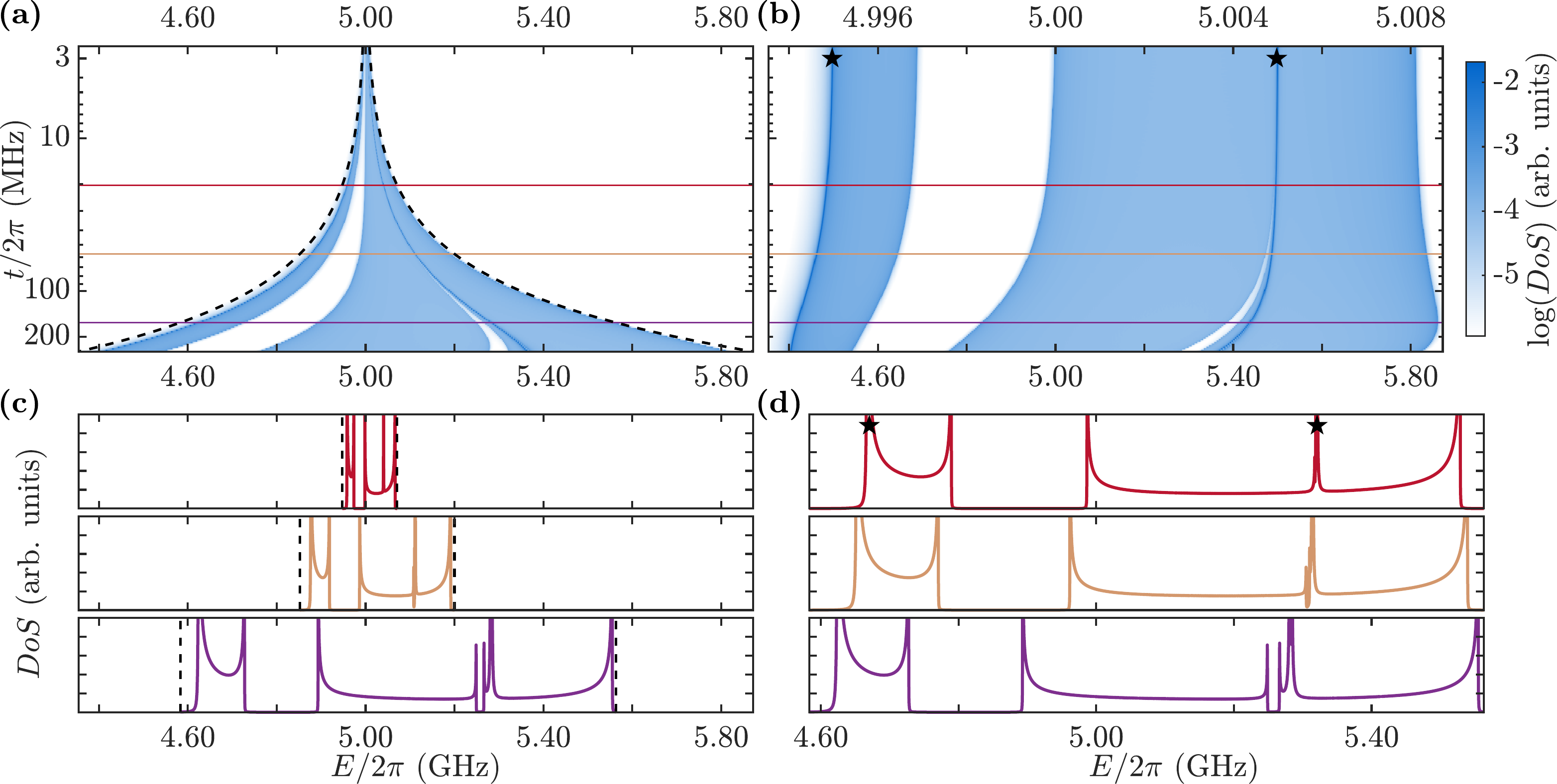}
    \caption{\textit{Simulated Density of States of the Half-Wave Modes versus Nearest-Neighbor Hopping $t$.} 
    Density of states (DoS) as a function of energy $E$ and hopping $t$ (a) in a fixed energy window for all $t$ on a logarithmic scale and (b) in an energy window that scales linearly with $t$. The energy range used in (b) is indicated in (a) by dashed black lines. (c)-(d) The line cuts of (a) and (b), respectively, are shown at three representative values of $t$ (colored horizontal lines in the corresponding 2D maps). Flat bands are marked by black stars in (b) and (d). In the simulations, we set $M = 100$.}
    \label{fig:HW_tuning}
\end{figure*}

The transmon-resonator coupling term in Eq.~\eqref{eq:lattice_Ham} can be written in terms of the creation and annihilation operators of the resonator and transmon modes as
\begin{nalign}
	\mathcal H_{\rm coupling} & = - C_c' \sum_{i,i'=1, j,j'=1}^{N, M} \left( \frac{1}{\sqrt{\mathbf L}} \mathbf V^{\rm T} \right)_{(i, j), (i', j')} \dot \theta_{i', j'} \dot \chi_i \\
	& = \frac{C_c'}{2} \sum_{i,i'=1, j,j'=1}^{N, M} \sqrt{ \frac{\omega_{i', j'} \Omega_i}{d_{i', j'} (C_q+C_c')} }\left( \frac{1}{\sqrt{\mathbf L}} \mathbf V^{\rm T} \right)_{(i, j), (i', j')} \left( b_{i', j'}^{\dagger} - b_{i', j'} \right) \left( a_i^{\dagger} - a_i \right)\\
	& \equiv \sum_{i=1}^N \sum_{i'=1, j'=1}^{N, M} g_{i; i', j'} \left( b_{i', j'}^{\dagger} - b_{i', j'} \right) \left( a_i^{\dagger} - a_i \right).
\end{nalign}
Here, the coupling strength between transmon $i$ and resonator-lattice mode $i', j'$ (with flux variable $\theta_{i', j'}$ and creation operator $b_{i', j'}^{\dagger}$) is defined as
\begin{equation}
    g_{i; i', j'} = \sum_{j=1}^M \frac{C_c'}{2} \sqrt{ \frac{\omega_{i', j'}^3 \Omega_i}{(C_q+C_c')} }\left( \frac{1}{\sqrt{\mathbf L}} \mathbf V^{\rm T} \right)_{(i, j), (i', j')}.
\end{equation}

\begin{figure}[t]
    \centering
    \includegraphics[width=0.95\linewidth]{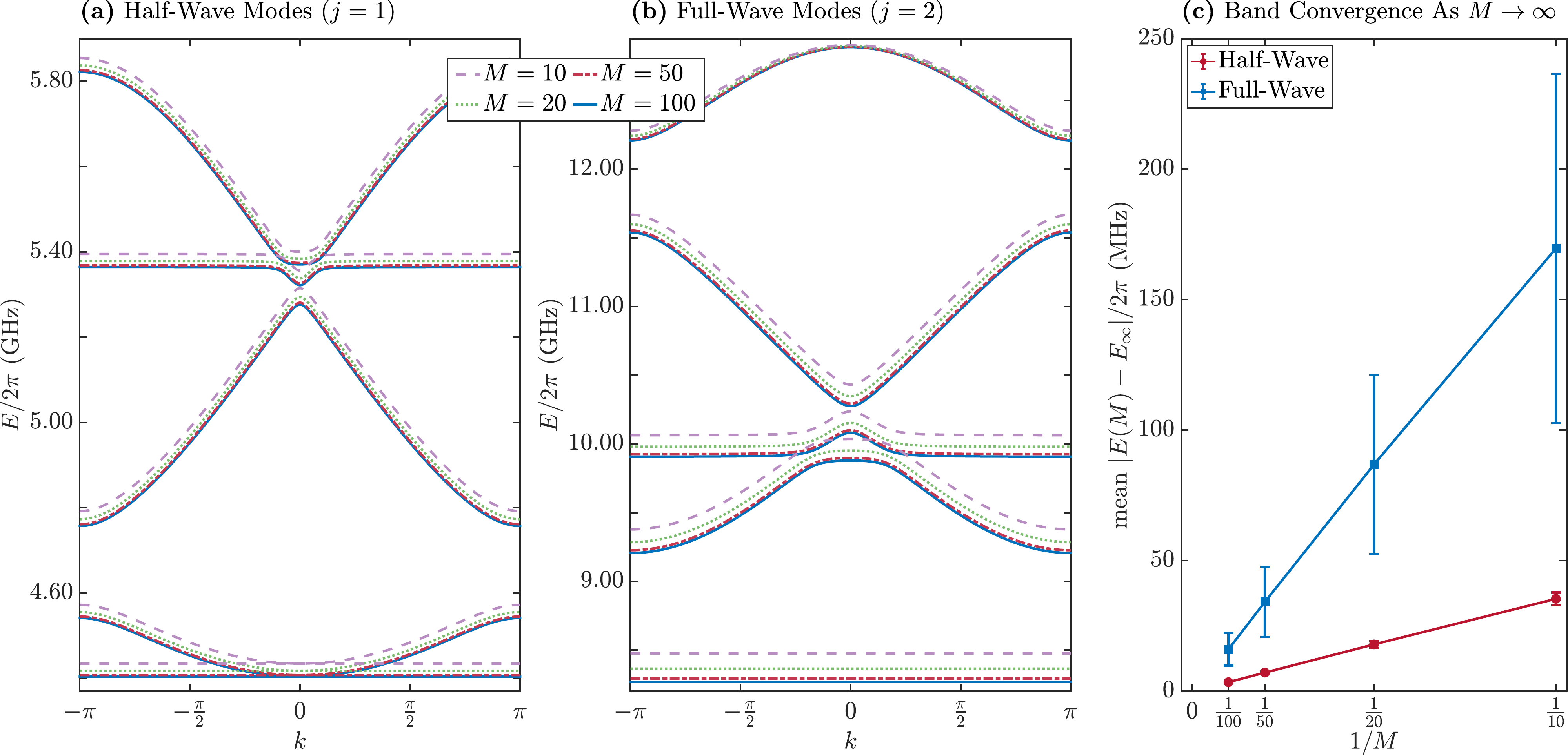}
    \caption{\textit{Band Structure Convergence with Increasing Mode Cutoff.} Band structure of a resonator-lattice with periodic boundary conditions [Eq.~\eqref{eq:res_lattice}] as a function of momentum $k$ for  various mode cutoffs $M$ for (a) half-wave and (b) full-wave modes. We use hopping $t=2 \pi \times 250$~MHz. (c) Mean absolute difference from the extrapolated band energies for $M=\infty$, averaged over all $k$ and all six bands for half-wave and full-wave modes as a function of $1/M$ along with the standard deviation from the mean.}
    \label{fig:band_structure_convergence}
\end{figure}

\vspace{0.1in}
The spectra shown in Figs.~\ref{fig:FW_tuning}~and~\ref{fig:HW_tuning} (corresponding to the full-wave and half-wave modes) are generated numerically by diagonalizing the resonator-lattice Lagrangian in Eq.~\eqref{eq:res_lattice} to obtain the normal lattice modes as in Eq.~\eqref{eq:res_lattice_diagonal}. These are the unperturbed resonator-lattice modes when the transmons are far detuned. 

The tight-binding single-mode Hamiltonian with hopping $t$~\cite{Kollar:2019linegraph, Koch:AnnPhysBerl, Koch:TRS_breaking, Houck_Nature_QS, Nunnenkamp:hopping, latticepaper} is a valid approximation of Eq.~\eqref{eq:res_lattice} for weak couplings, i.e., $C_c \ll C_0$. In this regime, we can ignore the terms coupling different intrinsic modes and write the resonator-lattice Lagrangian as
\begin{equation}
    \mathcal L_\text{res-lattice} \simeq \sum_{j=1}^M \mathcal L_\text{res-lattice}^{(j)},
\end{equation}
where
\begin{equation}
    \mathcal L_\text{res-lattice}^{(j)} = \sum_{i=1}^N \left( \frac{C_0+4C_c+C_c'}{2} \dot \phi_{i, j}^2 - \frac{j^2}{2L_0} \phi_{i, j}^2 \right) - C_c \sum_{\left< i, i' \right>} \sigma_{i, i'}^{(j)} \dot \phi_{i, j} \dot \phi_{i', j} \equiv \frac{1}{2} \dot \Phi_j^\text{T} \left( \tilde C_0 \mathbf I_N - C_c \mathbf\Sigma^{(j)} \right) \dot \Phi_j - \frac{j^2}{2L_0} \Phi_j^\text{T} \mathbf I_N \Phi_j.
\end{equation}
Here, $\Phi_j^\text{T} = (\phi_{1, j}, \phi_{2, j}, \dots, \phi_{N, j})$, $\tilde C_0 = C_0+4C_c+C_c'$ is the renormalized single-resonator capacitance, and ${\mathbf \Sigma^{(j)} = \left( \sigma_{i, i'}^{(j)} \right)_{N \times N}}$ includes the resonator connectivity information as well as possible negative hopping signs for odd-$j$ modes. Specifically, $\sigma_{i,i'}^{(j)} = (-1)^{j P_i^{i'} P_{i'}^i}$ for coupled resonators $i$ and $i'$, and $\sigma_{i,i'} = 0$ otherwise. Now, we can take the Legendre transform to obtain the single-mode Hamiltonian:
\begin{equation}\label{eq:Ham_j}
    \mathcal H_\text{res-lattice}^{(j)} = \frac{1}{2} \mathbf\Pi_j^\text{T} \left( \tilde C_0 \mathbf I_N - C_c \mathbf\Sigma^{(j)} \right)^{-1} \mathbf \Pi_j + \frac{j^2}{2L_0} \Phi_j^\text{T} \mathbf I_N \Phi_j \simeq \frac{1}{2\tilde C_0} \mathbf \Pi_j^\text{T} \left( \mathbf I_N + \frac{C_c}{\tilde C_0} \mathbf\Sigma^{(j)} \right) \mathbf \Pi_j + \frac{j^2}{2L_0} \Phi_j^\text{T} \mathbf I_N \Phi_j,
\end{equation}
with $\mathbf\Pi_j = \left( \tilde C_0 \mathbf I_N - C_c\mathbf\Sigma^{(j)} \right) \dot \Phi_j$ being the variable conjugate to $\Phi_j$. Equation~\eqref{eq:Ham_j} is mostly diagonal in terms of the $\pi_{i, j}$ (elements of $\mathbf \Pi_j$) and $\phi_{i, j}$:
\begin{equation}
    \mathcal H_\text{res-lattice}^{(j)} \simeq \sum_{i=1}^N \left( \frac{\pi_{i, j}^2}{2\tilde C_0} + \frac{j^2\phi_{i,j}^2}{2L_0} \right) + \frac{C_c}{\tilde C_0^2} \sum_{\left< i, i' \right>} \sigma_{i, i'}^{(j)} \pi_{i, j} \pi_{i', j},
\end{equation}
and in terms of the local creation operators $l_{i, j}^\dagger$ (not to be confused with the $b_{i, j}^\dagger$, the creation operators of the resonator-lattice normal modes in Eq.~\eqref{eq:res_lattice_Ham}):
\begin{nalign}
    \mathcal H_\text{res-lattice}^{(j)} & \simeq \sum_{i=1}^N j \tilde \omega_0 \left( l_{i, j}^\dagger l_{i, j} + \frac{1}{2} \right) - \frac{C_c}{\tilde C_0^2} \frac{\tilde C_0 j \tilde \omega_0}{2} \sum_{\left< i, i' \right>} \sigma_{i, i'}^{(j)} (l_{i, j}^\dagger - l_{i, j}) (l_{i', j}^\dagger - l_{i', j}) \\
    & \simeq \sum_{i=1}^N j \tilde \omega_0 \left( l_{i, j}^\dagger l_{i, j} + \frac{1}{2} \right) + j \frac{C_c}{2 \tilde C_0} \tilde \omega_0 \sum_{\left< i, i' \right>} \sigma_{i, i'}^{(j)} (l_{i, j}^\dagger l_{i', j} + l_{i, j} l_{i', j}^\dagger).
\end{nalign}
Here, $\tilde \omega_0 = 1/\sqrt{\tilde C_0 L_0}$ is the renormalized fundamental resonator frequency. In the last line, we have performed a rotating-wave approximation by throwing away the counter-rotating terms. The final result for the single-mode Hamiltonian is a tight-binding approximation with nearest-neighbor hopping of modes $j$ as $t_j = j t$, where $t$ is related to the circuit parameters as
\begin{equation}\label{eq:hopping_def}
    t = \frac{C_c}{2\tilde C_0} \tilde\omega_0 = \frac{C_c}{2 \left( C_0+4C_c+C_c' \right)^{3/2} L_0^{1/2}}.
\end{equation}

The coupling capacitance between each resonator and its transmon qubit $C_c'$ is taken to be one-half of the inter-resonator coupling capacitance:
\begin{equation}
    C_c' = \frac{C_c}{2}.
\end{equation}

The mode cutoff $M=100$ is chosen such that the band structures show acceptable convergence for the largest value of the hopping $t$ used in the numerical calculations (see Fig.~\ref{fig:band_structure_convergence}). As seen in Fig.~\ref{fig:band_structure_convergence}(c), for this value of $M$, the full-wave modes for $t=2\pi\times 250$ MHz are within $2\pi\times (16 \pm 8)$ MHz of the extrapolated bands for $M=\infty$.

Our numerical calculations assume periodic boundary conditions, whereas the experimental device contains a finite number of unit cells with physical ends. The periodic-boundary-condition calculation therefore isolates the bulk band structure and produces a smoother DoS representative of the infinite lattice, while the experiment exhibits a discrete set of finite-size modes, possible boundary-induced distortions near the band edges, and small frequency shifts associated with the termination of the lattice. These differences are expected to modify the detailed mode distribution but not the overall bulk band features emphasized here.

\section{Four-Wave-Mixing Characterization}\label{app:wmappendixfull}
\begin{figure*}[h]
\centering
    \includegraphics[width=1\textwidth]{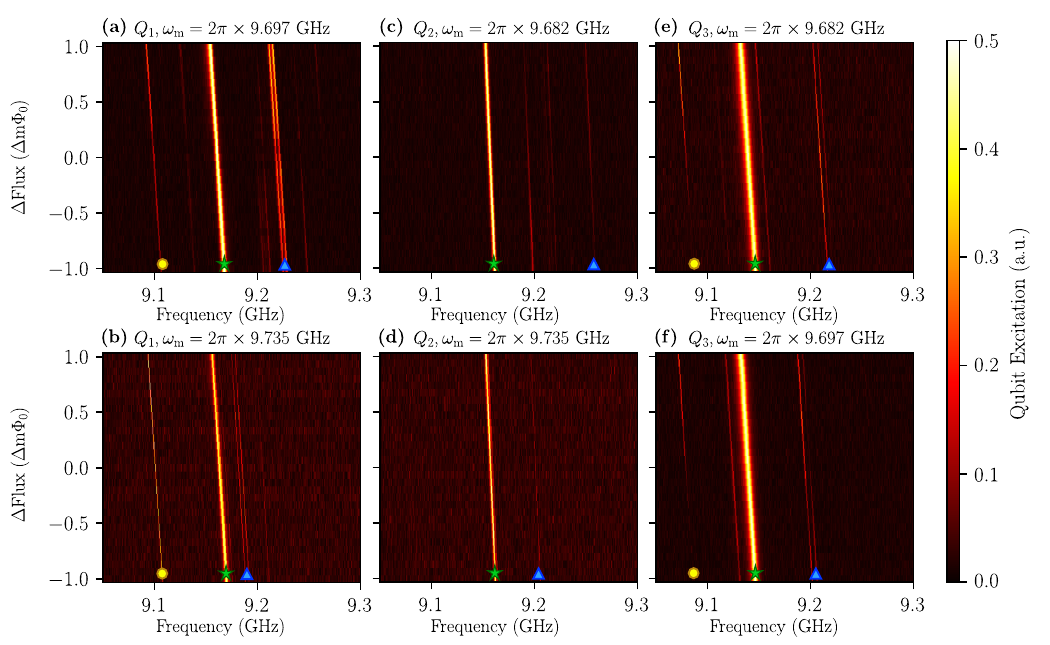}
    \vspace{-0.2cm}
    \caption{\label{fig:sweep} 
    \textit{Two-tone spectroscopy measurements versus applied flux and $\omega_m$.} 
    Each column contains data for one of the three qubits using two different monitor modes. The dominant spectral feature, marked with a green star, corresponds to the $g-e$ transition of the qubit. The $(g-f)/2$ transition, marked with a yellow circle, is sometimes visible. The four-wave-mixing resonance at $\wps$ is marked with a blue triangle. As $\wq$ is tuned via the applied flux, $\wps$ moves in the same direction with equal magnitude. 
    (a)-(b) Two-tone spectroscopy measurement of \sixteen using $\wm = 2\pi \times 9.697$ and $9.735$~GHz, respectively. The qubit transition frequencies are unaffected by the change in $\wm$, but the four-wave-mixing resonance $\wps$ is shifted \textit{down} by $\Delta \omega_m = 2\pi\times 38$~MHz.
    (c)-(d) Analogous measurement of \five using $\wm = 2\pi \times 9.682$ and $9.735$~GHz, respectively. 
    (e)-(f) Analogous measurement of \eight using $\wm = 2\pi \times 9.682$ and $9.697$~GHz, respectively.
    The data presented here verify the frequency matching condition in Eq.~\eqref{eq:wm} for all three qubits, and (a) contains the raw data corresponding to Fig.~\ref{fig:wmconcept}(a) in the main text.
    }
\end{figure*}

In this section, we describe the measurement protocol and the full characterization of the four-wave-mixing process. Qubit states in circuit QED are commonly measured using dispersive readout, which we employ here to probe the four-wave-mixing process. In the simplest case, a two-level-system qubit coupled to a single lattice mode is described by the Jaynes–Cummings Hamiltonian $H_\mathrm{JC}$ \cite{steckquantum},
\begin{equation}\label{eq:JC}
    H_\mathrm{JC} = \wm m^{\dagger} m - \frac{\wq}{2} \sigma_\mathrm{z} + g(m \sigma^{+} + m^{\dagger} \sigma^{-}),
\end{equation}
where $\wm$ is the bare resonator frequency, $\wq$ is the qubit transition frequency, and $g$ is the qubit–mode coupling strength. The first term represents the resonator Hamiltonian, with ladder operators $\hat m^{\dagger}$ and $\hat m$. The second term describes the qubit Hamiltonian as an effective spin-1/2 system, with the Pauli matrices or raising and lowering operators $\hat \sigma^{+}$ and $\hat \sigma^{-}$. The third term captures the interaction between the qubit and the resonator. In the dispersive regime, defined by $|\Delta| = |\wq - \wm| \gg g$, the Jaynes-Cummings Hamiltonian can be approximated as
\begin{equation}\label{eq:dispersive}
    H \simeq \left( \wm - \frac{g^2}{\Delta} \sigma_{z} \right) m^{\dagger} m + \left( \wq - \frac{g^2}{\Delta} \right) \frac{\sigma_{z}}{2}.
\end{equation}
In this regime, photons are not exchanged between the qubit and the resonator. Instead, the photon number and qubit state are conserved while their frequencies are shifted. Dispersive readout exploits this by detecting the qubit-state–dependent shift of the resonator frequency, $\wm^{\pm} = \wm \pm g^2/\Delta$, enabling qubit-state measurement \cite{Blais:CavityQED, wallraff:dispersiveReadout, Blais:CircuitQED, quantumengineersguide}. 

This dispersive readout scheme is implemented experimentally using two-tone spectroscopy, a form of pump–probe measurement, to determine the state of a qubit. A monitor tone, or a probe tone, is typically set to the frequency of the resonator corresponding to the qubit ground state. When an additional pump tone, $\omega_\mathrm{p}$, is swept across the qubit transition frequency $\wq$, qubit excitation can be induced, leading to a dispersive shift and a corresponding change in the monitor transmission or reflection (transitions labeled with a green star in Fig.~\ref{fig:sweep}). Similarly, the pump is tuned to the two-photon $(g-f)/2$ transition, excitations to the $\ket{f}$ are induced, which also shift the  monitor transmission or reflection. For a transmon qubit, this two-photon transition typically appears at a lower frequency than the $g-e$ transition due to the negative anharmonicity (transitions labeled with a yellow circle in Fig.~\ref{fig:sweep}).

As shown in Fig.~\ref{fig:wmcharacterization} of the main text, two-tone spectroscopy also reveals additional qubit excitations beyond the standard spectroscopy peaks, indicated with blue triangles in Fig.~\ref{fig:sweep}. These features are attributed to four-wave mixing, where a monitor photon at $\wm$ and a pump photon at $\wps$ are converted into a qubit photon at $\wq$ and a four-wave-mixing output photon at $\ww$.

In the main text, we showed that the four-wave-mixing process for \sixteen obeys Eq.~\eqref{eq:wm}. Raw data for those measurements are shown in Fig.~\ref{fig:sweep}, along with corresponding data for \five and \eightns, which obey the same energy conservation relation. When $\wq$ is tuned via flux biasing, $\wps$ shifts linearly with equal magnitude in the same direction as $\wq$. In contrast, when $\wm$ is varied, $\wps$ shifts with equal magnitude and the opposite sign from $\wm$. Together, these relations confirm energy conservation in which a single pump photon and a single monitor photon are absorbed, accompanied by the emission of a single qubit photon, along with a single output photon at $\ww$. Below in Appendix~\ref{app:wmappendixpowerscan}, we further verify the absorption of a single pump photon by examining the power saturation behavior of the wave-mixing process. The observed frequency conversion and the power saturation behavior show that the four-wave-mixing process is consistent with the theoretical model presented in Appendix~\ref{app:wminttheory}.

From measurements of the wave-mixing resonances in Fig.~\ref{fig:sweep}, the frequency of the output photon can be determined using the energy-conservation relation. The data for \sixteen in Fig.~\ref{fig:sweep}(a)–(b) and for \eight in Fig.~\ref{fig:sweep}(e)–(f) show the four-wave-mixing process at $\wps$ (blue triangle), corresponding to the upper flat band at $\ww = 2\pi \times 9.758$~GHz. This feature demonstrates strong coupling between the upper flat band and the off-axis qubits, \sixteen and \eightns. The data for \five in Fig.~\ref{fig:sweep}(c)–(d) shows four-wave mixing with the undriven lattice mode at $\ww = 2\pi \times 9.779$~GHz. As discussed in the main text and Appendix~\ref{app:quasi-1Dlattice}, \fivens, which is the on-axis qubit, does not couple to the upper flat band; only off-axis qubits do. The mode at $9.779$~GHz belongs to the strongly-coupled modes in the trivial dispersive bands.

\section{Excitation and Saturation versus Drive Power}\label{app:wmappendixpowerscan}
\begin{figure*}
\centering
    \includegraphics[width=1\textwidth]{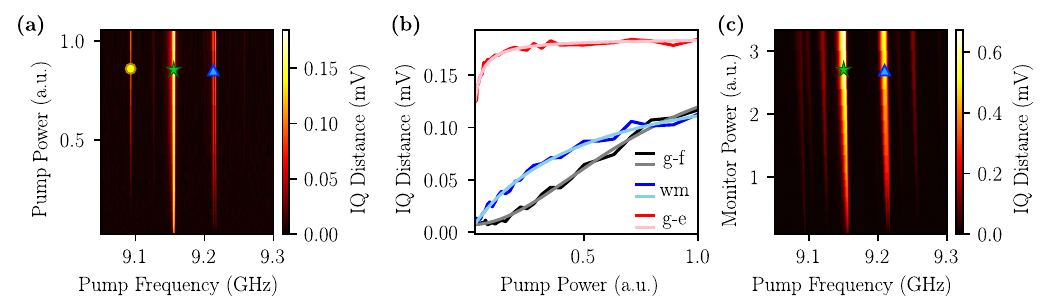}
    \vspace{-0.2cm}
    \caption{\label{fig:powersweep} 
    \textit{Pump Power Saturation and AC Stark Shift of the Four-Wave-Mixing Process.}
    (a) \sixteen is driven at fixed monitor and qubit frequencies while sweeping the pump power. Three spectral responses are observed, corresponding to the $(g-f)/2$ transition (yellow circle), the $g-e$ transition (green star), and the four-wave-mixing process (blue circle). The IQ distance provides a quantitative measure of steady-state population, with its limiting value corresponding to a maximally mixed state. 
    (b) Each curve shows the steady-state population as a function of pump power for different transitions. Both the $g-e$ transition and the four-wave-mixing process exhibit single-photon saturation as a function of pump power, whereas the $(g-f)/2$ transition displays two-photon saturation. At low power, single-photon saturation scales linearly, in contrast to the quadratic scaling of two-photon saturation. 
    (c) \sixteen is driven at fixed qubit frequency and pump power with $9.697$~GHz monitor mode. Sweeping the monitor power reveals two strong responses corresponding to the $g-e$ transition and the four-wave-mixing process. AC Stark shifts are observed for both transitions as the monitor power increases, resulting in negative frequency shifts.
    }
\end{figure*}

In the main text and Appendix~\ref{app:wmappendixfull}, we used energy conservation to establish that the observed four-wave-mixing process involves the absorption of a monitor photon and a pump photon, along with the excitation of a single qubit photon and the production of a single wave-mixing output photon. Here, we present further evidence from the power-saturation behavior of the four-wave-mixing process with respect to the pump power, showing that the nonlinear process is first-order with respect to the pump power.

We begin by reviewing the standard equation of power saturation in a two-level system. For a resonantly driven single-photon transition, the steady-state population $P_{ee}$ of the excited state $\ket{e}$ is given by~\cite{cohentanoudji}
\begin{equation}\label{eq:esat}
    P_{ee} = A \frac{B\mathcal{E}_{\mathrm{p}}^2}{1 + B\mathcal{E}_{\mathrm{p}}^2},
\end{equation}
where $\mathcal{E}_{\mathrm{p}}$ is the Rabi rate associated with the pump tone. For a conventional Rabi drive, Eq.~\eqref{eq:esat} takes the form $A = 1/2$ and $B = 2/\gamma_{e}^2$, where $\gamma_{e}$ denotes the decay rate of $\ket{e}$.

Here, we show that the power-saturation behavior of the four-wave-mixing process follows the form of a single-photon transition, shown in Eq.~\eqref{eq:esat}. Microscopically, the four-wave-mixing process arises from the effective Hamiltonian term $\mathcal{E}_{\mathrm{p}} \dqb ^\dag \fbop ^\dag \mop$ (see Appendix~\ref{app:wminttheory} for details) and is driven by the simultaneous absorption of a single pump photon at $\wps$, with Rabi rate $\mathcal{E}_\mathrm{p}$, and a monitor photon at fixed frequency $\wm$, with Rabi rate $\mathcal{E}_\mathrm{m}$. For fixed monitor power, the effective Rabi rate $\mathcal{E}_\mathrm{eff}$ associated with the four-wave-mixing process is proportional to $\mathcal{E}_\mathrm{p}$: $\mathcal{E}_\mathrm{eff} = C\mathcal{E}_\mathrm{p}$, where $C$ is a coefficient related to the monitor power and the monitor–pump frequency detuning. The coefficients of the steady-state population $P_{ee}^\mathrm{(wm)}$ of the four-wave-mixing process are $A=1/2$ and $B=2C^2/\gamma_\mathrm{e}^2$.

The same form can also describe the steady-state population $P_{ff}$ of a second excited state $\ket{f}$ when the pump tone induces a two-photon Raman transition. In this case, $\mathcal{E}_\mathrm{p}$ is replaced by the effective Raman Rabi rate $\mathcal{E}_\mathrm{Raman} = \sqrt{2}\mathcal{E}_\mathrm{p}^2/(2|\Delta_\mathrm{qp}|)$, where the qubit–pump detuning $|\Delta_\mathrm{qp}|$ for the Raman transition is $\alpha/2$ (half the anharmonicity of the transmon). The steady-state population $P_{ff}$ is given by
\begin{equation}\label{eq:fsat}
    P_{ff} \approx \tilde{A}' \frac{\tilde{B}'\mathcal{E}_{\mathrm{Raman}}^2}{1 + \tilde{B}'\mathcal{E}_{\mathrm{Raman}}^2} \approx \tilde{A} \frac{\tilde{B}\mathcal{E}_{\mathrm{p}}^4}{1 + \tilde{B}\mathcal{E}_{\mathrm{p}}^4}.
\end{equation}
Since the transmon is an anharmonic ladder system with comparable decay rates in the $\ket{e}$ and $\ket{f}$ states, the Raman transition induces significant population shelving in the $\ket{e}$ state. Thus, the coefficient $\tilde{A}$ is less than $1/2$, and the coefficient $\tilde{B}$ is approximately $1/(\alpha^2 \gamma_{e}^2)$.

The single- and two-photon transitions described by Eqs.~\eqref{eq:esat} and \eqref{eq:fsat} behave qualitatively differently at low pump powers~\cite{cohentanoudji}. For the qubit $g-e$ transition and the wave-mixing process, governed by Eq.~\eqref{eq:esat}, $P_{ee} \propto \mathcal{E}_{\mathrm{p}}^2$, scaling linearly with the applied pump power. In contrast, for the $(g-f)/2$ transition, governed by Eq.~\eqref{eq:fsat}, $P_{ff} \propto \mathcal{E}_{\mathrm{p}}^4$ and thus scales quadratically with the applied pump power.

Next, we examine the steady-state qubit population as a function of pump power experimentally using the IQ distance \cite{quantumengineersguide,wallraff:dispersiveReadout,Blais:CavityQED}, defined as the voltage displacement in the in-phase (I) and quadrature (Q) components between the initial and final states. It is conveniently expressed in vector form as the distance between the initial response $S(\omega_\mathrm{p}) = I(\omega_\mathrm{p}) +iQ(\omega_\mathrm{p})$ and the final response $S'(\omega_\mathrm{p}) = I'(\omega_\mathrm{p}) + iQ'(\omega_\mathrm{p})$ in the complex plane. Because of the QND nature of dispersive readout in superconducting qubits~\cite{Blais:CavityQED}, for a given qubit transition, the IQ distance $|S'(\omega_\mathrm{p})-S(\omega_\mathrm{p})| = \sqrt{\Delta I(\omega_\mathrm{p})^2 + \Delta Q(\omega_\mathrm{p})^2}$ serves as a direct measure of qubit excitation. Here we compare the conventional single-photon transition, $g-e$, the two-photon transition, $(g-f)/2$, and the wave-mixing transition. IQ distances as a function of pump power and frequency for all three transitions are shown in Fig.~\ref{fig:powersweep}(a).

The peak IQ distance for each transition at fixed pump power reflects the resonant steady-state qubit population. The resonant steady-state populations of the $g-e$ transition and the four-wave-mixing process are well fit by a single-photon excitation model [Eq.~\eqref{eq:esat}], exhibiting quadratic scaling with $\mathcal{E}_{\mathrm{p}}$ (i.e., linear scaling with pump power) at low powers, as shown in Fig.~\ref{fig:powersweep}(b). In contrast, the $(g-f)/2$ transition exhibits the expected behavior for a two-photon excitation process [Eq.~\eqref{eq:fsat}], showing quartic scaling with $\mathcal{E}_{\mathrm{p}}$ (i.e., quadratic scaling with pump power) at low powers. The observed pump-power dependence confirms that the four-wave-mixing feature satisfies Eq.~\eqref{eq:esat} and corresponds to a process involving one drive photon per qubit excitation.

The four-wave-mixing process exhibits an AC Stark shift~\cite{Schuster:acstark, Gambetta:acstark} as the monitor power increases, as shown in Fig.~\ref{fig:powersweep}(c). Consistent with earlier observations of the four-wave-mixing response, the main $g-e$ transition appears at $\wq = 2\pi \times 9.16$~GHz, accompanied by a pump tone associated with the four-wave-mixing process at $\wps = 2\pi \times 9.22$~GHz. As the monitor power increases, the AC Stark effect induces a negative shift of $\wq$, evidenced by the displacement of the $9.16$~GHz resonance. Correspondingly, the pump-tone frequency exhibits an equal negative shift, consistent with the four-wave-mixing condition in Eq.~\eqref{eq:wm}, which is tied to the absorption of a qubit excitation at $\wq$.

\section{Four-Wave-Mixing Theory Modeling}\label{app:wminttheory}
In this section, we model the four-wave-mixing process \cite{Devoret:firstwm, YvonneGao:thesis, Schoelkopf:earlywm} in the quasi-1D resonator lattice with three transmon qubits. This section focuses on only two harmonic oscillator modes (a monitor mode $m$ and a wave-mixing mode $w$) and a single anharmonic mode $q$, out of multitude of modes present in the resonator-transmon lattice model discussed in Appendix~\ref{app:bandcalculation}. 

Consider a system described by the Hamiltonian $H_\mathrm{tot}$ consisting of a transmon qubit with Hamiltonian $H_\mathrm{q}$
and lowering operator $\dqb$, a monitor mode denoted by the lowering operator $\mop$, and an undriven cavity mode denoted by the lowering operator $\fbop$. The transmon is coupled to both the monitor and the undriven cavity mode, with interaction Hamiltonian denoted by $H _{\mathrm{int}}$. The Hamiltonian contributions that only involve the monitor mode and the undriven cavity mode are contained in $H_\mathrm{res}$. The total Hamiltonian of the system is
\begin{align} 
\label{neq:H.tot}
&  H _{\mathrm{tot}} 
=  H _{\mathrm{q}} + H _{\mathrm{res}} + H _{\mathrm{int}}  + H _{\mathrm{p}} 
,\quad  
H _{\mathrm{q}} 
=  \wq \dqb ^\dag \dqb -\frac{\alpha}{2} \dqb ^\dag \dqb ^\dag \dqb  \dqb 
,\quad  
H_\mathrm{res} = \fbfreq \fbop ^\dag \fbop   + \monfreq \mop ^\dag \mop 
, \\
\label{neq:H.int.p}
& H_{\mathrm{int}}   = \gqb (\dqb +\dqb ^\dag ) (\fbop  +\fbop  ^\dag )+ \gqa (\dqb +\dqb ^\dag ) (\mop + \mop ^\dag ) 
,  \quad  
H _{\mathrm{p}} = \mathcal{E} _{\mathrm{p} } e ^{- i \omega_\mathrm{p} t} \dqb ^\dag  + \mathrm{H.c.}
, 
\end{align}
where $\wq$, $\wm$, and $\ww$ are angular frequencies of the qubit, monitor mode, and undriven cavity mode, respectively. 
The interaction term, $ H_\mathrm{int}$, represents the interaction between the qubit and the two cavity modes with the coupling strengths $\gqa$ and $\gqb$. $H_\mathrm{p}$ denotes the Hamiltonian contribution due to the strong pump tone driving the qubit, with drive amplitude $\mathcal{E} _\mathrm{p}$ and frequency $\omega_\mathrm{p}$.

In measurements, we observe peaks in the monitor-mode response when the following resonance condition is met:
\begin{align}\label{neq:fwm.cond}
\monfreq + \wps = \wq + \ww. 
\end{align}
Here, $\wps$ denotes the pump frequency at which the four-wave-mixing process becomes resonant. 
This strongly suggests that the observed resonance is due to a four-wave-mixing process driven by the pump.
We start from the total Hamiltonian in Eq.~\eqref{neq:H.tot} and transform to the rotating frame with respect to the pump frequency $\wps$. We make the rotating-wave approximation by ignoring the counter-rotating (i.e., off-resonant) contributions in the interaction Hamiltonian $H _{\mathrm{int}}$. The rotating-frame Hamiltonian is then
\begin{align}
\label{seq:Hfwm.rwa}
H _{\mathrm{RWA}} = - \Delta_\mathrm{q} \dqb ^\dag \dqb -\frac{\alpha}{2} \dqb ^\dag \dqb ^\dag \dqb  \dqb 
- \fbdet \fbop ^\dag \fbop  - \mondet \mop ^\dag \mop + \left (
\gqb \dqb \fbop ^\dag + \gqa \dqb \mop ^\dag 
+ \mathcal{E} _{\mathrm{p} }  \dqb ^\dag + \mathrm{H.c.}
\right ) 
.
\end{align}
Here, the detunings $\Delta_{\mathrm{x}} $ of mode x are defined as $\Delta_{\mathrm{x}} = \wps - \omega_{\mathrm{x}}$.

\vspace{0.1in}
\textbf{Four-wave-mixing process with a two-level system.} For simplicity, we first discuss the emergence of four-wave mixing for when the qubit is genuinely a two-level system. The corresponding Hamiltonian is obtained by taking $|\alpha| \rightarrow \infty$ in Eq.~\eqref{seq:Hfwm.rwa}:
\begin{align}
H _{\mathrm{TLS;RWA}} = - \Delta_\mathrm{q} \frac{\sigma_z}{2} 
- \fbdet \fbop ^\dag \fbop  - \mondet \mop ^\dag \mop + \left (
\gqb \sigma_- \fbop ^\dag + \gqa \sigma_- \mop ^\dag 
+ \mathcal{E} _{\mathrm{p} }  \sigma_{+} + \mathrm{H.c.}
\right )
,
\end{align}
where $\sigma_z$ and $\sigma_{\pm}$ denote the Pauli $Z$ and the raising/lowering operator for the qubit, respectively. In this regime, we can no longer perform a perturbative treatment directly in $\alpha$, but we can first diagonalize the undriven Hamiltonian (which conserves the total excitation number), and then derive leading-order perturbative correction due to the drive term $ \mathcal{E} _{\mathrm{p} }  (\sigma_{+} + \mathrm{H.c.})$. To simplify notation, we denote by $\ket{ n_{\mathrm{q}}, n_{\mathrm{w}}, n_{\mathrm{m}} } _{\text{n}} $ the eigenstates of the undriven Hamiltonian
\begin{align}
H _{\mathrm{TLS;norm}} = - \Delta_\mathrm{q} \frac{\sigma_z}{2} 
- \fbdet \fbop ^\dag \fbop  - \mondet \mop ^\dag \mop + \left (
\gqb \sigma_- \fbop ^\dag + \gqa \sigma_- \mop ^\dag 
+ \mathrm{H.c.}
\right )
,
\end{align}
where $n_{\mathrm{q}}, n_{\mathrm{w}}, n_{\mathrm{m}}$ denote   normal-mode occupations. We are interested in the following drive-induced matrix overlap:
\begin{align}
\label{seq:fwm.tls.overlap}
{}_{\text{n}} \!  \braket{1_{\mathrm{q}}, 1_{\mathrm{w}}, 0_{\mathrm{m}} | \mathcal{E} _{\mathrm{p} }  (\sigma_{+} + \mathrm{H.c.}) | 0_{\mathrm{q}}, 0_{\mathrm{w}}, 1_{\mathrm{m}} } \!  {}_{\text{n}} 
.
\end{align}
In the regime where the interaction strengths are much weaker than the detunings between the relevant modes, i.e.
\begin{align}
|\gqb | \ll |\Delta_\mathrm{q} - \fbdet | , \quad 
|\gqa | \ll |\Delta_\mathrm{q} - \Delta_\mathrm{m} | 
,
\end{align}
we can directly relate the normal-mode basis states $\ket{ n_{\mathrm{q}}, n_{\mathrm{w}}, n_{\mathrm{m}} } _{\text{n}} $ to the bare-mode basis states $\ket{ n_{\mathrm{q}}, n_{\mathrm{w}}, n_{\mathrm{m}} } $ via second-order perturbation theory:
\begin{nalign}
& \ket{0_{\mathrm{q}}, 0_{\mathrm{w}}, 1_{\mathrm{m}}} _{\text{n}}  
\simeq \ket{0_{\mathrm{q}}, 0_{\mathrm{w}}, 1_{\mathrm{m}}} 
- \frac{\gqa }{\qbmondet} \ket{1_{\mathrm{q}}, 0_{\mathrm{w}}, 0_{\mathrm{m}}} 
+ \frac{\gqb \gqa  }{ \Delta_{\text{qm}} \Delta_{\text{wm}}} \ket{0_{\mathrm{q}}, 1_{\mathrm{w}}, 0_{\mathrm{m}}} 
, \\
& \ket{1_{\mathrm{q}}, 1_{\mathrm{w}}, 0_{\mathrm{m}}} _{\text{n}}  
\simeq \ket{1_{\mathrm{q}}, 1_{\mathrm{w}}, 0_{\mathrm{m}}} + \frac{\gqa }{\qbmondet} \ket{0_{\mathrm{q}}, 1_{\mathrm{w}}, 1_{\mathrm{m}}} + \frac{\sqrt{2}g_\text{qw}}{\Delta_\text{qw}} \ket{0_\text{q}, 2_\text{w}, 0_\text{m}}
+\frac{\gqb \gqa }{ \Delta_{\text{qm}} \Delta_{\text{wm}}} 
\ket{1_{\mathrm{q}}, 0_{\mathrm{w}}, 1_{\mathrm{m}}} 
. \label{seq:TLS.normal.to.bare}
\end{nalign}
Here, we have defined relative detunings $\Delta_{\mathrm{xy}} $ of mode $y$ relative to $x$ via $\Delta_{\mathrm{xy}} = \omega_{\mathrm{x}} - \omega_{\mathrm{y}}$ to simplify notation. Substituting the above equations into the overlap in Eq.~\eqref{seq:fwm.tls.overlap}, we obtain
\begin{align}\label{seq:TLS.M.total}
{}_{\text{n}} \! \braket{1_{\mathrm{q}}, 1_{\mathrm{w}}, 0_{\mathrm{m}} | \mathcal{E} _{\mathrm{p} }  (\sigma_{+} + \mathrm{H.c.}) | 0_{\mathrm{q}}, 0_{\mathrm{w}}, 1_{\mathrm{m}} } \!  {}_{\text{n}} 
\simeq 2 \mathcal{E} _{\mathrm{p} } \frac{\gqb \gqa }{ \qbmondet \Delta_{\text{wm}}} 
.
\end{align}
This matrix overlap of a perturbative drive term describes the strength of a four-wave mixing process we would expect to be generated by a truly two-level system.

\vspace{0.1in}
\textbf{Four-wave-mixing process with a transmon qubit.} 
Now we can go back to the full transmon Hamiltonian within the rotating-wave approximation in Eq.~\eqref{seq:Hfwm.rwa}:
\begin{align}
H_{\mathrm{tm;RWA}}
=
-\Delta_{\mathrm{q}}\, \dqb^{\dagger}\dqb
-\frac{\alpha}{2}\,\dqb^{\dagger}\dqb^{\dagger}\dqb\dqb
-\fbdet\,\fbop^{\dagger}\fbop
-\mondet\,\mop^{\dagger}\mop
+\left(
\gqb\,\dqb\,\fbop^{\dagger}
+\gqa\,\dqb\,\mop^{\dagger}
+\mathcal{E}_{\mathrm{p}}\,\dqb^{\dagger}
+\mathrm{H.c.}
\right).
\end{align}
To lowest nontrivial order in the four-wave-mixing amplitude of interest,
it suffices to keep transmon levels $\ket{0}_{\mathrm{q}},\ket{1}_{\mathrm{q}},\ket{2}_{\mathrm{q}}$,
so that
\begin{align}
\dqb \simeq \ket{0}_{\mathrm{q}}\!\bra{1}_{\mathrm{q}}
+\sqrt{2}\,\ket{1}_{\mathrm{q}}\!\bra{2}_{\mathrm{q}} .
\label{seq:dqb.trunc}
\end{align}

As in the two-level system discussion, we first diagonalize the undriven Hamiltonian:
\begin{align}
H_{\mathrm{tm;norm}}
&=
-\Delta_{\mathrm{q}}\, \dqb^{\dagger}\dqb
-\frac{\alpha}{2}\,\dqb^{\dagger}\dqb^{\dagger}\dqb\dqb
-\fbdet\,\fbop^{\dagger}\fbop
-\mondet\,\mop^{\dagger}\mop
+\left(
\gqb\,\dqb\,\fbop^{\dagger}
+\gqa\,\dqb\,\mop^{\dagger}
+\mathrm{H.c.}
\right),
\end{align}
and denote its normal-mode eigenstates by $\ket{n_{\mathrm{q}},n_{\mathrm{w}},n_{\mathrm{m}}}_{\mathrm{n}}$.
We then evaluate the drive-induced overlap
\begin{align}
\label{seq:fwm.tr.overlap}
{}_{\mathrm{n}}\!\braket{1_{\mathrm{q}},1_{\mathrm{w}},0_{\mathrm{m}}
\big|
\mathcal{E}_{\mathrm{p}}(\dqb^{\dagger}+\dqb)
\big|
0_{\mathrm{q}},0_{\mathrm{w}},1_{\mathrm{m}}}_{\mathrm{n}}.
\end{align}

In the dispersive regime
\begin{align}
|\gqb|\ll|\Delta_{\mathrm{q}}-\fbdet|,
\qquad
|\gqa|\ll|\Delta_{\mathrm{q}}-\mondet|,
\end{align}
we can express the normal-mode states perturbatively in terms of the bare states
$\ket{n_{\mathrm{q}},n_{\mathrm{w}},n_{\mathrm{m}}}$.
Keeping only the components that contribute at order
$\mathcal{O}(\gqb\gqa)$, we write
\begin{align}
\ket{0_{\mathrm{q}},0_{\mathrm{w}},1_{\mathrm{m}}}_{\mathrm{n}}
&\simeq
\ket{0_{\mathrm{q}},0_{\mathrm{w}},1_{\mathrm{m}}}
-\frac{\gqa}{\qbmondet}\ket{1_{\mathrm{q}},0_{\mathrm{w}},0_{\mathrm{m}}}
+ \frac{\gqb \gqa  }{ \Delta_{\text{qm}} \Delta_{\text{wm}}}\ket{0_{\mathrm{q}},1_{\mathrm{w}},0_{\mathrm{m}}},
\label{seq:tr.normal.to.bare.i}
\\
\ket{1_{\mathrm{q}},1_{\mathrm{w}},0_{\mathrm{m}}}_{\mathrm{n}}
&\simeq
\ket{1_{\mathrm{q}},1_{\mathrm{w}},0_{\mathrm{m}}}
+\frac{\gqa}{\qbmondet}\ket{0_{\mathrm{q}},1_{\mathrm{w}},1_{\mathrm{m}}} + \frac{\sqrt{2}g_\text{qw}}{\Delta_\text{qw}} \ket{0_\text{q}, 2_\text{w}, 0_\text{m}} \nonumber \\
&+ \left( \frac{\gqb \gqa  }{ \Delta_{\text{qm}} \Delta_{\text{wm}}} - \frac{2 g_{\text{qw}} g_{\text{qm}}}{(\Delta_{\text{qw}}-\alpha) \Delta_\text{wm}} \right)\ket{1_{\mathrm{q}},0_{\mathrm{w}},1_{\mathrm{m}}}
-\frac{\sqrt{2}\,\gqb}{\Delta_{\text{qw}}-\alpha}\ket{2_{\mathrm{q}},0_{\mathrm{w}},0_{\mathrm{m}}}.
\label{seq:tr.normal.to.bare.f}
\end{align}
There are new contributions in Eq.~\eqref{seq:tr.normal.to.bare.f} compared to the two-level case that arise because $\fbop$ can exchange an excitation with the transmon on the
$\ket{1}_{\mathrm{q}}\leftrightarrow\ket{2}_{\mathrm{q}}$ transition, which carries
a $\sqrt{2}$ matrix element [cf.~Eq.~\eqref{seq:dqb.trunc}] and a shifted energy denominator
$\qbfbdet-\alpha$ (i.e., the detuning is displaced by the anharmonicity for the $1\!\to\!2$ manifold).

Substituting Eqs.~\eqref{seq:tr.normal.to.bare.i}~and~\eqref{seq:tr.normal.to.bare.f}
into Eq.~\eqref{seq:fwm.tr.overlap}, we find that 
the transmon matrix overlap is
\begin{nalign}
\label{seq:tr.M.total}
{}_{\mathrm{n}}\!\braket{1_{\mathrm{q}},1_{\mathrm{w}},0_{\mathrm{m}}
\big|
\mathcal{E}_{\mathrm{p}}(\dqb^{\dagger}+\dqb)
\big|
0_{\mathrm{q}},0_{\mathrm{w}},1_{\mathrm{m}}}_{\mathrm{n}}
&=
2\mathcal{E}_{\text p} \frac{\gqb\gqa}{\qbmondet\,\Delta_{\text{wm}}}
\left(
1
-\frac{\Delta_{\text{qw}}}{\Delta_{\text{qw}}-\alpha}
\right)
\\
&=
-2\,\mathcal{E}_{\mathrm{p}}\,
\frac{\gqb\gqa\,\alpha}{\qbmondet\,\Delta_{\text{wm}}\,(\Delta_{\text{qw}}-\alpha)}.
\end{nalign}
Note that this overlap vanishes for $\alpha \rightarrow 0$ (i.e., when the transmon is a harmonic oscillator mode), and reproduces the two-level system result in Eq.~\eqref{seq:TLS.M.total} for $|\alpha| \rightarrow \infty$. Additionally, in the regime when the two-photon resonance condition holds,
\begin{align}
\monfreq + \wps - \wq = \ww \simeq \wq - \alpha 
\Rightarrow 
\Delta_\text{qw} - \alpha \simeq 0,
\end{align}
the two states $\ket{1_{\mathrm{q}}, 1_{\mathrm{w}}, 0_{\mathrm{m}}} _{\text{n}} $ and $\ket{2_{\mathrm{q}}, 0_{\mathrm{w}}, 0_{\mathrm{m}}} _{\text{n}} $ become near resonant, and the perturbative analysis above is no longer valid. 

The four-wave-mixing drive-induced matrix overlap \textit{with a transmon} [Eq.~\eqref{seq:tr.M.total}] differs from that \textit{with a two-level system} [Eq.~\eqref{seq:TLS.M.total}] by a multiplicative factor $-\alpha/(\Delta_\text{qw}-\alpha)$. In the experiments reported in the main text, this factor is approximately $0.17$; see, for example, the values given in the caption of Fig.~\ref{fig:wmconcept}. In this limit, where $\alpha$ is small compared to $\Delta_\text{qw}$, the wave-mixing process can also be described using the techniques of nonlinear optics, as shown below.

\vspace{0.1in}
\textbf{Four-wave mixing-process with a transmon qubit in the perturbative regime (small $\alpha$).} 
Finally, we provide an analysis to show how the presented results arise as a four-wave mixing process for small nonlinearity $\alpha$ by applying the general machinery of nonlinear optics. We assume the weak drive regime, where $| \mathcal{E}_\mathrm{p} | \ll | \Delta_\mathrm{q} | $, so that we can apply a displacement transformation on all the involved bosonic modes: 
\begin{align}
\dqb \to \delta \dqb + \langle \dqb \rangle_\mathrm{p} , \quad 
\fbop \to \delta \fbop + \langle \fbop \rangle_\mathrm{p} , \quad  
\mop \to \delta \mop  + \langle \mop \rangle_\mathrm{p} . \quad 
\end{align}
In the weak-drive regime, we expect $\langle \dqb \rangle_\mathrm{p} \simeq  \mathcal{E}_\mathrm{p}/\Delta_\mathrm{q} $; the higher-order corrections to the expectation values are not crucial to the current discussion, and we omit them for conciseness.
Keeping only the leading-order (in $\alpha$ and $|\langle \dqb \rangle_\mathrm{p} |$) contributions,  
we obtain  
\begin{align}
H_{\mathrm{RWA}} \simeq 
& - (\Delta_\mathrm{q} + 2 \alpha |\langle \dqb \rangle_\mathrm{p}| ^{2} ) \delta \dqb ^\dag \delta \dqb 
- \frac{\alpha}{2} (\langle \dqb \rangle_\mathrm{p} ^{2} \delta \dqb ^{\dag 2} +  \mathrm{H.c.} )
- \fbdet \delta \fbop ^\dag \delta \fbop  - \Delta_\mathrm{m} \delta \mop ^\dag \delta \mop 
\nonumber \\
& + \left (- \alpha \langle \dqb \rangle_\mathrm{p} 
\delta \dqb ^\dag \delta  \dqb  ^\dag\delta \dqb  + 
\gqb \delta \dqb \delta \fbop ^\dag + \gqa \delta \dqb \delta \mop ^\dag  + \mathrm{H.c.}
\right ) 
. 
\end{align}
We consider the weak-drive limit so that $|\Delta_\mathrm{q}| \gg \alpha |\langle \dqb \rangle_\mathrm{p}| ^{2}$ and the quadratic contributions to the transmon dynamics are negligible compared to the detuning term. We omit the quadratic contributions henceforth for simplicity: 
\begin{align}
\label{neq:H.rwa.disp}
H_{\mathrm{RWA}} \simeq 
& - \Delta_\mathrm{q} \delta \dqb ^\dag \delta \dqb 
- \fbdet \delta \fbop ^\dag \delta \fbop  - \Delta_\mathrm{m} \delta \mop ^\dag \delta \mop 
+ \left (- \alpha \langle \dqb \rangle_\mathrm{p} 
\delta \dqb ^\dag \delta  \dqb  ^\dag\delta \dqb  + 
\gqb \delta \dqb \delta \fbop ^\dag + \gqa \delta \dqb \delta \mop ^\dag  + \mathrm{H.c.}
\right ) 
.
\end{align}
We diagonalize the quadratic part of the Hamiltonian Eq.~\eqref{neq:H.rwa.disp} with a beamsplitter-type unitary transformation. Moreover, in the regime where
the interaction strengths are much weaker than the detunings between the relevant modes, i.e.
\begin{align}
|\gqb | \ll |\Delta_\mathrm{q} - \fbdet | , \quad 
|\gqa | \ll |\Delta_\mathrm{q} - \Delta_\mathrm{m} | 
,
\end{align}
this procedure can be done perturbatively in coupling strengths $|\gqb | $ and $|\gqa | $ as 
\begin{nalign}
\label{seq:fwm.normal.to.bare}
& \delta \dqb  \to \delta \dqb ' +  \frac{\gqb }{\fbdet -\Delta_\mathrm{q}} \delta \fbop ' + \frac{\gqa }{\mondet -\Delta_\mathrm{q}} \delta \mop '
= \delta \dqb ' +  \frac{\gqb }{ \qbfbdet} \delta \fbop ' + \frac{\gqa }{\qbmondet} \delta \mop ', \\
& \delta \fbop \to \delta \fbop ' - \frac{\gqb }{\fbdet -\Delta_\mathrm{q}} \delta \dqb ' 
= \delta \fbop ' - \frac{\gqb }{ \qbfbdet} \delta \dqb ' , \\
& \delta \mop \to \delta \mop ' - \frac{\gqa }{\mondet -\Delta_\mathrm{q}} \delta \dqb '
= \delta \mop ' - \frac{\gqa }{\qbmondet} \delta \dqb ',
\end{nalign}
which formally corresponds to a Schrieffer-Wolff transformation (only on the quadratic part of the Hamiltonian). 

The nonlinearity in Eq.~\eqref{neq:H.rwa.disp} now becomes
\begin{align}
&H '_{\mathrm{FWM}} = - \left (\alpha \langle \dqb \rangle_\mathrm{p} 
\delta \dqb ^\dag \delta  \dqb ^\dag \delta \dqb + \mathrm{H.c.} \right )
\nonumber \\
=& - \alpha \langle \dqb \rangle_\mathrm{p} \left ( {\delta \dqb '} ^\dag +  \frac{\gqb }{ \qbfbdet} {\delta \fbop '} ^\dag + \frac{\gqa }{\qbmondet} {\delta \mop ' } ^\dag \right )
\left ( {\delta \dqb '} ^\dag +  \frac{\gqb }{ \qbfbdet} {\delta \fbop '} ^\dag + \frac{\gqa }{\qbmondet} {\delta \mop ' } ^\dag \right )
\left ( \delta \dqb ' +  \frac{\gqb }{ \qbfbdet} \delta \fbop ' + \frac{\gqa }{\qbmondet} \delta \mop ' \right )
 + \mathrm{H.c.} 
\nonumber \\
=& - \left ( 2 \alpha \langle \dqb \rangle_\mathrm{p} \frac{\gqb }{ \qbfbdet} \frac{\gqa }{\qbmondet} 
 {\delta \dqb '} ^\dag {\delta \fbop '} ^\dag \delta \mop ' + \mathrm{H.c.} \right )+\ldots 
,
\end{align}
where we keep the relevant four-wave-mixing terms that will become resonant at the frequencies probed in experiments. 
Specifically, we now go to the rotating frame defined by the diagonal terms in $H _{\mathrm{RWA}}$ in Eq.~\eqref{neq:H.rwa.disp}. The first term becomes $- \left (2 \alpha \langle \dqb \rangle_\mathrm{p} \frac{\gqb}{\qbfbdet} \frac{\gqa}{\qbmondet} 
e ^{i(\Delta_\mathrm{q} + \fbdet - \mondet) t} {\delta \dqb '} ^\dag {\delta \fbop '} ^\dag \delta \mop '+ \mathrm{H.c.} \right ) $.
We see that the four-wave-mixing term in the displaced rotating frame becomes resonant when the detunings satisfy the condition
\begin{align}
\Delta_\mathrm{q} + \fbdet = \mondet 
, 
\end{align}
which matches Eq.~\eqref{neq:fwm.cond}. Thus, at the four-wave-mixing resonance, the drive amplitude can be written as 
\begin{align}
\langle \dqb \rangle_\mathrm{p} = \frac{\mathcal{E}_\mathrm{p}}{\Delta_\mathrm{q}}
= \frac{\mathcal{E}_\mathrm{p}}{\mondet -\fbdet } =  \frac{\mathcal{E}_\mathrm{p}}{\Delta_\mathrm{wm} }
,
\end{align}
and the perturbative Hamiltonian becomes (retaining only the resonant term)
\begin{align}
&H '_{\mathrm{FWM}} \propto - 2 \alpha \frac{\mathcal{E}_\mathrm{p}}{\Delta_\mathrm{wm} } \frac{\gqb }{ \qbfbdet} \frac{\gqa }{\qbmondet} 
{\delta \dqb '} ^\dag {\delta \fbop '} ^\dag \delta \mop ' + \mathrm{H.c.}
 ,
\end{align}
which agrees with Eq.~\eqref{seq:tr.M.total} when the anharmonicity is much smaller than the detuning, $\alpha \ll |\Delta_{\text{qw}}|$.

\end{document}